\documentclass[11pt]{article}
\usepackage{graphicx}
\usepackage{amsmath}
\usepackage{amssymb}
\newcommand{\eS}{{\epsilon_S}}
\newcommand{\eT}{{\epsilon_T}}
\newcommand{\eP}{{\epsilon_P}}
\newcommand{\eR}{{\epsilon_R}}
\newcommand{\eL}{{\epsilon_L}}
\newcommand{\eLc}{{\epsilon_L^{(c)}}}
\newcommand{\eLv}{{\epsilon_L^{(v)}}}
\newcommand{\teS}{{\tilde{\epsilon}_S}}
\newcommand{\teT}{{\tilde{\epsilon}_T}}
\newcommand{\teP}{{\tilde{\epsilon}_P}}
\newcommand{\teL}{{\tilde{\epsilon}_L}}
\newcommand{\teR}{{\tilde{\epsilon}_R}}
\newcommand{\beq}{\begin{equation}}
\newcommand{\eeq}{\end{equation}}
\newcommand{\beqa}{\begin{eqnarray}}
\newcommand{\eeqa}{\end{eqnarray}}
\newcommand{\beqan}{\begin{eqnarray*}}
\newcommand{\eeqan}{\end{eqnarray*}}
\newcommand{\ba}{\begin{array}}
\newcommand{\ea}{\end{array}}
\newcommand{\bea}{\begin{eqnarray}}
\newcommand{\eea}{\end{eqnarray}}

\newcommand{\jr}{}
\renewcommand{\thefootnote}{\alph{footnote}}

\begin{document}
%%%%%%%%%%%%%%%%%%%%%%%%%%%%%%%%%%%%%%%%%%%%%%%%%%%%%%%%%%%%%%%%%%%%%%%%%%%%%%%%%%%%%%%%%%%%
\begin{flushright}
NPAC-13-03
%September  2012
\end{flushright}
\vspace{2.5cm}

\begin{center}
{\bf \LARGE Prospects for precision measurements in nuclear $\beta$ decay at the LHC era}

\bigskip\bigskip
\large
O. Naviliat-Cuncic$^{1,}$\footnote{E-mail:~\textsf{naviliat@nscl.msu.edu}}
and
M. Gonz\'alez-Alonso$^{2,}$\footnote{E-mail:~\textsf{gonzalezalon@wisc.edu}}

\bigskip\bigskip
$^1${\small\it NSCL and Department of Physics and
Astronomy, Michigan State University, MI 48824, USA}\\
$^2${\small\it Department of Physics, University of Wisconsin-Madison,
WI 53706, USA}
\end{center}

\normalsize\rm
\renewcommand{\thefootnote}{\arabic{footnote}}
%%%%%%%%%%%%%%%%%%%%%%%%%%%%%%%%%%%%%%%%%%%%%%%%%%%%%%%%%%%%%%%%%%%%%%%%%%%%%%%%%%%%%%%%%%%%
\begin{abstract}
Precision measurements in nuclear $\beta$ decay offer a sensitive
window to search for new physics beyond the standard electroweak model and
allow also the determination of the fundamental weak vector coupling
in processes involving the lightest quarks.
Searches for new physics are also a strong motivation for experiments carried out at
the high energy frontier reached at the most powerful particle colliders.
It is instructive to confront
results from the low energy and the high energy frontiers
in order to look for possible complementarities and orient new avenues for
experiments at low energies.
We review here the status of constraints on new physics obtained from
nuclear and neutron decays and compare them to those from other
semi-leptonic processes and from the LHC. We stress the requirements of new
precision experiments
in $\beta$ decay in order to impact the search for new physics
at the light of current and projected LHC results.
We describe recent experimental results and ongoing developments in nuclear
and neutron $\beta$ decay, with emphasis
on their planned goals to improve present limits on exotic weak couplings.
\end{abstract}
%%%%%%%%%%%%%%%%%%%%%%%%%%%%%%%%%%%%%%%%%%%%%%%%%%%%%%%%%%%%%%%%%%%%%%%%%%%%%%%%%%%%%%%%%%%%
%%%%%%%%%%%%%%%%%%%%%%%%%%%%%%%%%%%%%%%%%%%%%%%%%%%%%%%%%%%%%%%%%%%%%%%%%%%%%%%%%%%%%%%%%%%%
\section{Introduction}

Searches for physics beyond the Standard electroweak Model (SM) 
are carried out both
at the high energy frontier, attained at the most powerful particle
colliders, as well as at the high precision frontier, looking for 
deviations from SM predictions
in low background environments, where high sensitivities to small effects
can often be achieved.

Nuclear $\beta$ decay and neutron decay have played a crucial role in
the development of the ``$V-A$'' theory of the weak interaction, which
was eventually embedded in the wider framework of the SM
\cite{weinberg09,severijns06}. 
Today, one of the main motivations for improving the experimental
sensitivities of
precision experiments in nuclear and neutron decays is
the search for possible non-SM or ``exotic'' interactions that would manifest
themselves
through genuine scalar or tensor terms in semi-leptonic weak processes.

The tests of the SM and the searches for New Physics (NP) in nuclear and
neutron decays have been the
subject of several recent reviews, with focus either on
experiments in nuclear $\beta$ decay
\cite{severijns06,towner10,severijns11,severijns13}
or on experiments using cold or ultra-cold neutrons
\cite{abele08,nico09,dubbers11}.

A recurrent question addressed to precision measurements carried out
at low energies concerns their sensitivity to NP as compared with
results obtained at high energy, and currently at the Large Hadron
Collider (LHC).

In this paper we review first the status of constraints on scalar and tensor
couplings obtained from precision measurements in
nuclear and neutron decays. 
We stress the observation that, for the most precise measurements,
the only relevant parameter providing stringent constraints on exotic interactions
is the Fierz interference term, through its contribution to other
correlation coefficients.

We confront next the most precise results obtained in $\beta$ decay with constraints
obtained from other
semi-leptonic processes and also with results from the LHC\cite{Bhattacharya:2011qm,Cirigliano:2012ab,Cirigliano2013},
paying special attention to those observables
that are linear on the exotic couplings. The use of an Effective Field
Theory (EFT) framework allows us to bridge through the low energy and
the high energy searches and to compare their sensitivities to new physics.

Considerable experimental effort using neutrons and nuclei are underway
worldwide with the aim to improve the precision on decay observables.
We describe recent experimental results and current developments in nuclear
and neutron decays, and discuss their
precision goals and sensitivities to exotic weak couplings.

%%%%%%%%%%%%%%%%%%%%%%%%%%%%%%%%%%%%%%%%%%%%%%%%%%%%%%%%%%%%%%%%%%%%%%%%%%%%%%%%%%%%%%%%%%%%
\section{Theoretical description}

In the SM, semi-leptonic processes at the quark-lepton level are described
by the exchange of the charged vector bosons, $W^\pm$. Since the mass of the bosons
are significantly larger than the energies involved in nuclear and neutron
$\beta$ decays, the interaction Lagrangian for these processes
takes the usual $(V-A)\times(V-A)$ form

\bea
{\cal L}_{\rm SM}  &=&
- \frac{G_F V_{ud}}{\sqrt{2}} \ 
\bar{e}  \gamma_\mu  (1 - \gamma_5)   \nu_e  \cdot \bar{u}   \gamma^\mu  (1 - \gamma_5) d ~,
\label{eq:lSM-lowE}
\eea
where $G_F$ is the Fermi coupling and $V_{ud}$ is the element of the
Cabibbo-Kobayashi-Maskawa (CKM) matrix involved in the weak coupling of the
lightest quarks.
This Lagrangian provides the framework for the calculations of observables to
leading order, that will be compared with experimental results.
However, due to the precision of current experiments, the calculation of
SM predictions requires to take into account corrections to this contact
form arising from the
finiteness of the $W$ mass and from electroweak radiative
corrections \cite{Czarnecki:2004cw,Ando:2004rk}.

For the inclusion of
NP effects in nuclear and neutron $\beta$ decays,
it is very useful to follow an EFT approach. This model-independent framework
allows us to compare the sensitivity of these processes with other low-energy
charged-current observables and also with measurements carried out at high energy colliders.

%%%%%%%%%%%%%%%%%%%%%%%%%%%%%%%%%%%%%%%%%%%%%%%%%%%%%%%%%%%%%%%%%%%%%%%%%%%%%%%%%%%%%%%%%%%%
\subsection{Quark-level Effective Lagrangian}
\label{sec:quark-level-Leff}

Assuming that the particles not included in the SM are much heavier than the
energy scales relevant for nuclear and neutron $\beta$
decay, they can again be integrated out along with the $W$ boson and the rest
of heavy SM particles. The low-scale $O(1 \ {\rm GeV})$ effective Lagrangian for semi-leptonic transitions is then given by~\cite{Bhattacharya:2011qm,Cirigliano:2012ab}\footnote{For the sake of simplicity we do not considered operators involving $\nu_\mu$ or $\nu_\tau$. The generalization is straightforward and the general formulae can be found in Ref.~\cite{Cirigliano:2012ab}.}.
\begin{eqnarray}
{\cal L}_{\rm eff}  &=&
- \frac{G_F V_{ud}}{\sqrt{2}} \left[ \ \left( 1 +  \eL \right) \
\bar{e}  \gamma_\mu  (1 - \gamma_5)   \nu_e  \cdot \bar{u}   \gamma^\mu  (1 - \gamma_5)  d \right.
\nonumber \\
& + &
\teL  \ \ \bar{e}  \gamma_\mu  (1 + \gamma_5)   \nu_e  \cdot \bar{u}   \gamma^\mu  (1 - \gamma_5)  d
\nonumber\\
&+&   \eR   \  \   \bar{e}  \gamma_\mu  (1 - \gamma_5)   \nu_e
\cdot \bar{u}   \gamma^\mu  (1 + \gamma_5)  d
\nonumber\\
&+&
\tilde{ \epsilon}_R   \  \   \bar{e}  \gamma_\mu  (1 +  \gamma_5)   \nu_e
\cdot \bar{u}   \gamma^\mu  (1 + \gamma_5)  d
\nonumber\\
&+&  \eS  \  \  \bar{e}  (1 - \gamma_5) \nu_e  \cdot  \bar{u} d
 \ + \  \teS  \  \  \bar{e}  (1 +  \gamma_5) \nu_e  \cdot  \bar{u} d
 \nonumber \\
 &-& \eP  \  \   \bar{e}  (1 - \gamma_5) \nu_e  \cdot  \bar{u} \gamma_5 d
 \ -  \  \teP  \  \   \bar{e}  (1 + \gamma_5) \nu_e  \cdot  \bar{u} \gamma_5 d
 \nonumber \\
 &+&
\eT    \   \bar{e}   \sigma_{\mu \nu} (1 - \gamma_5) \nu_e    \cdot  \bar{u}   \sigma^{\mu \nu} (1 - \gamma_5) d
\nonumber\\
&+&
\left.
\teT      \   \bar{e}   \sigma_{\mu \nu} (1 + \gamma_5) \nu_e    \cdot  \bar{u}
 \sigma^{\mu \nu} (1 + \gamma_5) d \right] + {\rm h.c.}~.
\label{eq:leff-lowE}
\end{eqnarray}
%
%\end{subequations}
%where $e,u,d$ denote the electron, up- and down-quark mass eigenfields, while $\nu_e$ represents the electron-neutrino.
The $\epsilon_i$ and $\tilde{\epsilon}_i$ complex coefficients are functions of the masses and couplings of the new particles, in the same way that the Fermi constant $G_F$ is a function of the weak coupling and the $W$ mass.
The specific  expressions of these coefficients within the minimal supersymmetric standard model
can be found in e.g. Ref.~\cite{Profumo:2006yu}.

For the sake of generality we have included right-handed (RH) neutrinos in the low-energy particle content, but they can easily be removed setting $\tilde{\epsilon}_{L,R,S,P,T}=0$. It is worth noticing that operators involving RH neutrinos contribute quadratically to the observables, what makes their effect on the experiments much smaller.

The effective Lagrangian in Eq.~(\ref{eq:leff-lowE}) describes the effect of NP not only in nuclear and neutron $\beta$
decay, but also in other processes like for example $\pi^{\pm} \to \pi^0 e^{\pm} \nu$. The details of the hadronization are obviously different, with different form factors needed in each process, but the underlying dynamics is the same.%, as shown with this effective Lagrangian.

After removing an overall phase, we have ten real couplings and nine phases that can be probed comparing precise low-energy experiments and accurate SM calculations. Given the smallness of these couplings it is useful to work at linear order in them to identify their main effect on the different observables. As explained above, in this approximation we can neglect the $\tilde{\epsilon}_i$ couplings, since they involve RH neutrinos. Moreover, in nuclear and neutron decays the pseudo-scalar coupling $\eP$ can also be neglected since the associated hadronic bilinear vanishes in the non-relativistic approximation. In this approximation the low-energy effective Lagrangian can be written as
\begin{eqnarray}
{\cal L}_{\rm eff}  &=&
- \frac{G_F V_{ud}}{\sqrt{2}} \
\left[ 1 + \mbox{Re} \left(  \eL+\eR  \right) \right] \times
\label{eq:leff-lowE-linear} \\
&&\times \left\{
\bar{e}  \gamma_\mu (1 - \gamma_5) \nu_{e}  \cdot \bar{u}
\gamma^\mu \left[ 1 - \left(1-2\eR \right) \gamma_5 \right]   d \right.
\nonumber  \\
&&+~  \eS  \  \  \bar{e}  (1 - \gamma_5) \nu_{e}  \cdot  \bar{u} d
%\nonumber \\&&
% -~ \eP  \  \   \bar{e}  (1 - \gamma_5) \nu_{e}  \cdot  \bar{u} \gamma_5 d
 \nonumber \\
 &&+~
\left.
\eT    \   \bar{e}   \sigma_{\mu \nu} (1 - \gamma_5) \nu_{e}    \cdot  \bar{u}
\sigma^{\mu \nu} (1 - \gamma_5) d \right\}
+{\rm h.c.}~,
\nonumber
\end{eqnarray}
where an overall phase has been omitted.

Furthermore, in neutron decay it is usual to extract the axial-vector form factor $g_A$ (or the so-called mixing ratio in the case of nuclear decays) from experiments. The extracted quantity contains an unobservable NP contribution \cite{Bhattacharya:2011qm,Herczeg2001vk}
\bea
g_A \to g_A~\mbox{Re}\left[ \frac{1+\eL-\eR}{1+\eL+\eR} \right]
\approx g_A \left[ 1 - 2 \mbox{Re}(\eR) \right] + {\cal O}\left( \epsilon_i^2\right)~,
\eea
that can only be probed if a precise lattice calculation of $g_A$ becomes available.

All in all, we see that there are six couplings left in this approximation:
\begin{itemize}
\item The real part of $\eL+\eR$ that produces a shift in the overall normalization and can be absorbed
in a redefinition of $V_{ud}$, with the only consequence being the violation of the unitarity condition 
of the first raw of the CKM quark mixing matrix,
$|V_{ud}|^2+|V_{us}|^2+|V_{ub}|^2\neq 1$.
\item The real parts of the scalar $\eS$ and tensor $\eT$ couplings that modify the energy distributions and CP-even correlation coefficients.
\item The imaginary parts of the axial-vector $\eR$, scalar $\eS$ and tensor $\eT$ couplings that modify CP-odd correlation coefficients.
\end{itemize}
These six couplings represent the only linear NP effects in nuclear and neutron $\beta$ decay. Consequently we expect strong bounds on them, whereas weak bounds are expected to be obtained for the rest of the couplings.

%%%%%%%%%%%%%%%%%%%%%%%%%%%%%%%%%%%%%%%%%%%%%%%%%%%%%%%%%%%%%%%%%%%%%%%%%%%%%%%%%%%%%%%%%%%%
\subsection{Nucleon-level effective couplings}
\label{sec:nucl-Eff-Coup}

The next step in the theoretical description is matching the quark-level effective Lagrangian, Eq.~(\ref{eq:leff-lowE}), onto a nucleon-level effective Lagrangian. Working at leading order in momentum transfer, the neutron-to-proton matrix elements can be written as
\bea
\langle p | \bar{u} \Gamma d | n \rangle = g_\Gamma  \, \bar{\psi}_p \Gamma \psi_n ~,
\eea
with $\Gamma = 1, \gamma_5, \gamma_\mu, \gamma_\mu \gamma_5, \sigma_{\mu \nu}$.
Proceedings in this way, the Lee-Yang effective Lagrangian is obtained
\cite{lee56,jackson57a}\footnote{Notice that the original paper of Lee and Yang \cite{lee56}
uses a different definition of $\gamma_5$ that we do not follow here.
The definitions of all the couplings, $C^{(\prime)}_i$ used here and in Refs.~\cite{lee56,jackson57a}
are however the same.
%%%
%Notice also that Ref.~\cite{Herczeg2001vk} defines the couplings $C_A, C'_{V,S,T}$ with an overall minus sign compared to ours.
%%%
}
\bea
- {\cal L}_{n \to p e^- \bar{\nu}_e}
&=& ~~\bar{p}~n ~\left( C_S \bar{e} \nu_e - C'_S \bar{e} \gamma_5 \nu_e  \right) \nonumber\\
&&+~ \bar{p}\gamma^\mu n \left( C_V \bar{e} \gamma_\mu \nu_e - C'_V \bar{e} \gamma_\mu \gamma_5 \nu_e  \right)  \nonumber\\
&&+~ \bar{p}\sigma^{\mu\nu} n \left( C_T \bar{e} \sigma_{\mu\nu_e} \nu_e - C'_T \bar{e} \sigma_{\mu\nu} \gamma_5 \nu_e  \right)  \nonumber\\
&&\mathbf{-}~ \bar{p}\gamma^\mu \gamma_5 n \left( C_A \bar{e} \gamma_\mu \gamma_5 \nu_e - C'_A \bar{e} \gamma_\mu \nu_e  \right)  \nonumber\\
&&+~ \bar{p} \gamma_5 n ~ \left( C_P \bar{e} \gamma_5 \nu_e - C'_P \bar{e} \nu_e  \right)+\mbox{h.c.}~.
\label{eq:leffJTW}
\eea
The effective couplings $C_i$, $C_i'$ ($i \in \{V,A,S,T\}$) can be expressed in terms of the parton-level parameters as
%
%\label{eq:matchLY}
\bea
C_{i} &=& \frac{G_F}{\sqrt{2}} \,  V_{ud} \,  \overline{C}_{i}    \\
\overline{C}_V &=& g_V  \left(1 + \eL + \eR
+ \teL + \teR
\right)  \\
\overline{C}_V' &=& g_V  \left(1 + \eL + \eR
- \teL - \teR
\right)  \\
\overline{C}_A &=& - g_A  \left(1 + \eL - \eR
- \teL   +  \teR
\right)  \\
\overline{C}_A' &=& - g_A  \left(1 + \eL - \eR
+ \teL   -  \teR
\right)  \\
\overline{C}_S &=&   g_S  \, \left(  \eS  + \teS \right) \\
\overline{C}_S'  &=&   g_S  \, \left(  \eS  -  \teS \right) \\
\overline{C}_P &=&   g_P  \, \left(  \eP  - \teP \right) \\
\overline{C}_P'  &=&   g_P  \, \left(  \eP  +  \teP  \right) \\
\overline{C}_T &=&    4 \, g_T \, \left( \eT   + \teT \right)  \\
\overline{C}_T '  &=&    4 \, g_T \, \left( \eT   -  \teT \right) ~.
\eea

The use of this Lagrangian to study NP effects is fully justified due to the small magnitude
of such effects.
On the other hand, in order to have a precise determination of the SM contributions one has,
again,
to take into account higher order terms in the momentum transfer expansion, like weak-magnetism, as well as other sub-leading corrections like electromagnetic effects \cite{Czarnecki:2004cw,Ando:2004rk,Gardner:2000nk}.
%Ref.~\cite{Ando:2004rk} studied neutron beta decay within an EFT approach showing in an elegant way how these NLO effects can be included in a model-independent way (except for a

Using these relations and the results of Ref.~\cite{jackson57b} one can work out the dependence of neutron and nuclear $\beta$ decay observables on the short-distance parameters $\epsilon_i$ and $\tilde{\epsilon}_i$. %As we mentioned above, the effect of the pseudo-scalar interactions associated to the couplings $C_{P,P'}$ is zero in the non-relativistic approximation and it can thus be neglected.

For the hadron-level coefficients, the results of Sec. \ref{sec:quark-level-Leff} indicate that the only quantities that can affect linearly the observables are (i) the real part of $(C_V+C_V')$ that shifts the value of $V_{ud}$, (ii) the real and imaginary parts of the combinations $(C_{S}+C'_{S})/C_V$ and $(C_T+C'_T)/C_A$ and (iii) the relative phase between $C_V$ and $C_A$.

The precise knowledge of both vector and axial-vector charges $g_{V,A}$ is needed to accurately calculate the SM contribution. The vector charge is $g_V = 1$ up to second-order isospin symmetry breaking corrections that can be safely neglected. The axial-vector charge $g_A$ cannot be accurately calculated from first principles, and the usual strategy is to keep it as an independent parameter that can be extracted from experiments with high precision.

The presence of exotic scalar and tensor interactions introduces two additional form factors $g_{S,T}$ that we need to know in order to convert the measured quantities into the quark-level parameters $\epsilon_{S,T}$. In the rest of this article we will use the recent lattice QCD determinations $g_S = 0.8 \pm 0.4$ and $g_T = 1.05 \pm 0.35$ in the $\overline{MS}$ scheme and at the renormalization scale $\mu=2$~GeV \cite{Bhattacharya:2011qm}. %, where the uncertainties include an estimate of all the systematic effects.
Notice that, given the smallness of the scalar and tensor couplings $\epsilon_{S,T}$, it is not necessary to have such a precise determination of $g_{S,T}$ as for the $g_{V,A}$ couplings.

The bounds derived for $\epsilon_{S,T}$ depend both on our ability to accurately calculate the $g_{S,T}$ form factors and to perform precise experiments that can put strong constraints on the hadronic couplings $C_i$.
The level of precision needed in the lattice determination so that the final bounds on $\epsilon_{S,T}$
would be dominated by experimental errors has been studied in Ref.~\cite{Bhattacharya:2011qm},
assuming a determination of the Fierz term $b$ (see below) at the $10^{-3}$ level in future neutron decay
experiments. That study concluded that an improvement by a factor of two in the current lattice error will be necessary\footnote{Given the R-fit method used in Ref.~\cite{Bhattacharya:2011qm}, the lower bounds of $g_{S,T}$ are actually a better indicator of the final $\epsilon_{S,T}$ bounds than the errors $\delta g_{S,T}$. It is thus technically more accurate to say that a future determination of $\epsilon_{S,T}$ from a measurement of $b_n$ at the level of $10^{-3}$ will be dominated by the experimental error as long as $g_S^{min} >0.64$ and $g_T^{min}>0.84$.}, what has motivated a renewed effort in the lattice QCD community to improve the present results \cite{Bhattacharya:2011qm, Green:2012ej}.%Gupta:2012az

%%%%%%%%%%%%%%%%%%%%%%%%%%%%%%%%%%%%%%%%%%%%%%%%%%%%%%%%%%%%%%%%%%%%%%%%%%%%%%%%%%%%%%%%%%%%
\subsection{Nuclear matrix elements}

An additional step is necessary for nuclear decays to connect the hadron-level effective Lagrangian, Eq.\eqref{eq:leffJTW}, with the nuclei involved in the transition. 

Working at leading order in the non-relativistic approximation, this step can be done in a simple way and only two new quantities are needed: (i) the Fermi nuclear matrix element, $M_F$, that encodes the nuclear effects in the vector- and scalar-mediated transitions, and (ii) the Gamow-Teller nuclear matrix element, $M_{GT}$, that plays the same role in transitions mediated by tensor and axial-vector interactions. In this same limit the pseudo-scalar nuclear matrix element vanishes. Moreover $M_F$ (and other terms in the momentum transfer expansion, like weak-magnetism) can be calculated exactly in the isospin symmetry limit.

Once again the use of this approximation is justified to study the effect of NP at the current level of precision, whereas sub-leading corrections have to be taken into account in the calculation of the SM contribution. This includes higher order terms in the momentum transfer expansion like weak-magnetism \cite{grenacs85}, electromagnetic effects or nuclear structure dependences \cite{Czarnecki:2004cw,holstein74,hardy09}. Forbidden effects in allowed transitions, associated with an orbital angular momentum of the lepton pair $\ell\neq0$, are in principle negligible at the current level of precision due to the $(qR)^\ell$ suppression, where $R$ is the nucleus radius. However if the allowed matrix
elements $M_{F}$ or $M_{GT}$ happen to be suppressed, the forbidden effects could become significant what could be the case for nuclear decays with large $ft$-values.

%%%%%%%%%%%%%%%%%%%%%%%%%%%%%%%%%%%%%%%%%%%%%%%%%%%%%%%%%%%%%%%%%%%%%%%%%%%%%%%%%%%%%%%%%%%%
\subsection{Weak-scale operator basis}

The effective Lagrangian given in Eq.\eqref{eq:leff-lowE} allows us to compare the NP sensitivity of different low-energy experiments, even when the hadrons involved are different, like pion decay and neutron decay. Under some reasonable assumptions such a comparison can be done also with experiments performed at much higher energies and with some processes involving neutral currents, due to the $SU(2)$ gauge invariance.

Indeed, if we assume that the new fields introduced by the theory that supersedes the SM are not only heavier than the scales relevant for $\beta$ decay experiments but also heavier than the energy scale of current collider experiments, we can then describe the physics at that scale also through an effective Lagrangian. This high-energy effective Lagrangian includes all SM fields as active degrees of freedom and has the following structure \cite{LEFF,Cirigliano:2012ab}
\bea
\label{eq:BW}
%{\cal L}_{eff} = {\cal L}_{SM} + \frac{1}{\Lambda^2} {\cal L}_6 + \ldots
{\cal L}_{eff} = {\cal L}_{SM} + \frac{1}{\Lambda^2} \sum_i \alpha_i {\cal O}_i^{(6)} + \ldots ~,
\eea
where ${\cal O}_i^{(6)}$ are $SU(2)_L\times U(1)_Y$-invariant dimension-six effective operators generated by the exchange of the new fields that we integrated out, $\alpha_i$ are the associated Wilson coefficients and $\Lambda$ is the characteristic NP scale. %The effective Lagrangian contains also higher-dimensional operators, but they are expected to be suppressed by powers of $v/\Lambda$.

It is possible to identify twelve effective operators %\footnote{Five of them involve right-handed neutrinos.}
that generate a tree-level contribution to nuclear and neutron $\beta$ decay, either through a modification of the $W$ boson vertex to fermions or introducing a new four-fermion interaction. It is then possible to relate the Wilson coefficients of both theories, i.e. $\epsilon_i = f (\alpha_j)$ and likewise for $\tilde{\epsilon}_i$. These matching conditions, which can be found in Ref.~\cite{Cirigliano:2012ab}, allow us to understand the implications of collider searches for low-energy experiments and vice versa.%, in a model-independent way.

%We note here that the individual effective couplings $\epsilon_{S,P,T}$ ($\tilde{\epsilon}_{S,P,T}$) and the corresponding  hadronic matrix elements can display a strong scale dependence. Throughout the paper, we will quote estimates and bounds for the $\epsilon_i$ ($\tilde{\epsilon}_i$) at the renormalization scale  $\mu=2$~GeV in the $\overline{\rm MS}$ scheme.

%%%%%%%%%%%%%%%%%%%%%%%%%%%%%%%%%%%%%%%%%%%%%%%%%%%%%%%%%%%%%%%%%%%%%%%%%%%%%%%%%%%%%%%%%%%e
\section{Correlations in allowed $\beta$ decay}
\label{sec:correlations}

The relations between the experimentally accessible angular and energy distributions and the hadronic couplings $C_i$ and $C^\prime_i$ was established in the seminal paper of Jackson, Treiman and Wyld (JTW) for allowed nuclear transitions \cite{jackson57a,jackson57b}. Their formalism is also valid for neutron decay when making the appropriate substitutions for the Fermi and Gamow-Teller matrix elements.

The decay rate distributions are expressed in terms of the total energies $E_i$ and the momenta ${\bf p}_i$ of the decay products, with $i = e, \nu$ for the $\beta$ particle and the neutrino respectively, and also the spins {\bf J} and \boldmath$\sigma$\unboldmath$_e$ of the decaying system and of the $\beta$ particle. Those distributions include correlation terms which are scalar, pseudo-scalar or mixed products of the kinematic vectors.
For example, the angular and energy distribution of the electron and neutrino in the decay of unpolarized nuclei or neutrons has the form \cite{jackson57b}
\begin{equation}
\label{eq:decay-rate}
\omega(E_e) = \omega_0(E_e) \xi \left( 1 + b \frac{m}{E_e} + a \frac{\bf{p}_e \cdot \bf{p}_\nu}{E_e E_\nu} \right) ~,
\end{equation}
where the function $\omega_0(E_e)$ includes the phase space factor and the Fermi function, $m$ is the electron mass and the coefficients $\xi$, $b$, and $a$ contain the dynamics of the decay. This includes the nuclear matrix elements, $M_F$ and $M_{GT}$, and the effective couplings $C_i$ and $C^\prime_i$ \cite{jackson57a,jackson57b}. Any specific experimental configuration fixes the magnitude of the various correlation terms and then the sensitivity to the coefficients, $a$, $b$, etc.

\begin{figure}[!hbt]
\centerline{
\includegraphics[width=0.35\columnwidth]{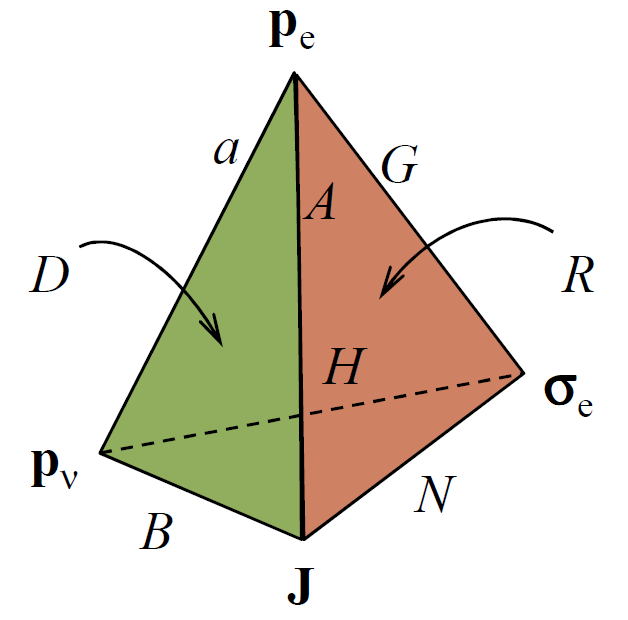}
}
\caption{
\label{fig:correlations}
Pictorial representation of the two-fold (edges) and three-fold (sides) correlation
coefficients between the kinematic vectors (vertices). Adapted from Ref.~\cite{dubbers11}.
}
\end{figure}

For a quick reference to the most common correlations coefficients, Fig.~\ref{fig:correlations} shows a pictorial representation, in the form of a tetrahedron. The vertices correspond to the kinematic vectors, the edges to the two-fold correlations and the faces
to the three-fold correlations.

At leading order in the approximations discussed above, the SM corresponds to $C^\prime_V/C_V = C^\prime_A/C_A = 1$, all other parameters being zero. 
Deviations of experimental results from the values of the coefficients predicted in the SM would provide an indication for new physics or a departure from the allowed approximation for nuclear decays.

It can easily be seen from the expressions of the coefficients \cite{jackson57b} that all of them receive
linear NP contributions which are not suppressed to order $\alpha$ except the coefficients $a$, $A$ and $G$. Among those, the coefficients that have been accessed experimentally, either directly or in conjunction with other coefficients are:
the Fierz interference term $b$, the neutrino asymmetry parameter $B$, the polarization-spin correlation $N$, the $\beta$ longitudinal polarization from
polarized nuclei $Q$ and the triple correlations $R$ and $D$. %The remaining coefficients are H, N, L, S, U and V.
These latter are sensitive to a possible relative phase between the couplings
arising from time-reversal violation.

Measurements of $N$ and $Q$ require the analysis of the spin of $\beta$ particles emitted from polarized neutrons or nuclei, what makes such experiments very demanding and challenging.

In most experimental conditions, the measured coefficients receive a linear contribution via the Fierz term of the form
\begin{equation}
\label{eq:Fierz-corr}
\tilde{X} = \frac{X} { 1 + b \langle m/E_e \rangle } ~,
\end{equation}
where $X = a, A, B$, etc. stands for any of the correlation coefficients and $\langle \, \rangle$ denotes the weighted average over the observed part of the $\beta$ energy spectrum.
Such a contribution arises for instance in measurements of the $\beta\nu$ angular correlation, $a$, which is quadratic on the exotic couplings. For $C^{(\prime)}_i \ll 1$, the sensitivity of $\tilde{a}$ to those couplings becomes then dominated by the contribution due to the Fierz term. 
Under such conditions, the factor $\langle m/E_e \rangle$ becomes an important quantity since it can suppress or enhance significantly the NP sensitivity. This factor depends on the transition and the details of the experiment, and is typically in the range $0.2-0.7$ \cite{severijns06}.

The contribution of the Fierz term can however adversely affect the sensitivity of other coefficients. For example, for pure Gamow-Teller transitions, which are sensitive to tensor contributions, the $b$ and $B$ coefficients have the form \cite{jackson57b}
\bea
\label{eq:bGT}
b_{GT} &=& \pm \gamma~ {\rm Re} \left( \frac{C_T + C^\prime_T}{C_A} \right)~,\\
\label{eq:BGT}
B_{GT} &=& \lambda_{J'J} \left[\pm 1 + \frac{\gamma m}{E_e}~
{\rm Re} \left( \frac{C_T + C^\prime_T}{C_A} \right) \right]~,
\eea
where $\lambda_{J'J}$ is a spin factor \cite{jackson57a}, $\gamma = \sqrt{1 - (\alpha Z)^2}$ with $Z$
the atomic number of the daughter nucleus and the upper (lower) sign refers to electron (positron) decay.
Following Eq.~(\ref{eq:Fierz-corr}) we have then
\begin{equation}
\label{eq:B-tilde}
\tilde{B}_{GT} \approx  B_{GT} \left( 1 - \frac{m}{E_e} b_{GT} \right) \approx \pm\lambda_{J'J}~.
\end{equation}
 
This shows that the measured coefficient looses then the linear sensitivity to tensor couplings for such transitions. A similar, albeit partial, suppression of sensitivity has been observed for $\tilde{B}$ in neutron decay~\cite{Bhattacharya:2011qm}. These suppressions are due to the fact that $B$ and $b$ have similar (linear) dependence on the exotic couplings.

The sensitivity of $B$ to the couplings actually depends on the initial spin of the transition, on the spin sequence and on the mixing ratio $\rho = C_A M_{GT}/(C_V M_F)$ between the Gamow-Teller and Fermi contributions. Thus the cancellation due to the contribution of $b$ to $\tilde{B}$ observed in pure Gamow-Teller processes does not necessarily happen in all mixed transitions.\footnote{It is certainly conceivable to perform simultaneous measurements of ratios of coefficients, like $B/a$, so that the linear sensitivity to the NP contribution in $B$ is not lost. However, such a measurement has
not yet been performed.}

In summary, the discussion above shows that the sensitivity of
correlation coefficients which are linear in the exotic couplings
can strongly be affected by the contribution of the Fierz term.
For coefficients that depend quadratically on the couplings,
the dominant sensitivity to these couplings will arise
from the contribution of the Fierz. 
For transitions with large end-point energies, the contribution
of the Fierz term can however be strongly weaken. 
We illustrate this quantitatively in the next sections.

%%%%%%%%%%%%%%%%%%%%%%%%%%%%%%%%%%%%%%%%%%%%%%%%%%%%%%%%%%%%%%%%%%%%%%%%%%%%%%%%%%%%%%%%%%%%
\section{Constraints from nuclear and neutron $\beta$ decay}
\label{sec:nuclear}

We review here the most stringent constraints on NP from the measurement
of energy and angular distributions in nuclear and neutron decays.

The use of the hadronic-level couplings $C^{(\prime)}_i$ of JTW is convenient at this stage because the comparison of sensitivity between different observables and measurements does not require the knowledge of the form factors $g_{V,A,S,T}$. However in the next section we will convert the most precise results to the quark-level notation $\epsilon_i$ and $\tilde{\epsilon}_i$ in order to be able to compare the NP sensitivity of these decays with other low- and high-energy observables.

Following the results of Sec.~\ref{sec:nucl-Eff-Coup} we present here the
constraints on quantities that can affect linearly the observables, namely:
(i) the real part of the vector couplings
$(C_V + C^\prime_V)$;
(ii) the real parts of the scalar and tensor couplings $(C_S + C^\prime_S)/C_V$
and $(C_T + C^\prime_T)/C_A$; and (iii) the imaginary parts of the scalar and
tensor couplings as well as the relative phase between $C_V$ and $C_A$.
%The extraction of the limits assumes real couplings for the $V$ and $A$ interactions as well as maximal parity violation for these couplings.

%%%%%%%%%%%%%%%%%%%%%%%%%%%%%%%%%%%%%%%%%%%%%%%%%%%%%%%%%%%%%%%%%%%%%%%%%%%%%%%%%%%%%%%%%%%%
\subsection{Limits on real vector couplings}
\label{sec:real-VA-nuclear}

The effect of the NP contribution to the real part of $(C_V + C^\prime_V)$ shifts the value of $V_{ud}$ and
can be probed through the unitarity test of the CKM matrix.

The most precise determination of $V_{ud}$ comes from the study of super-allowed
Fermi transitions \cite{hardy09,towner10}. Together with the recent
determinations of $V_{us}$ and $V_{ub}$ \cite{Moulson:2013wi}, it offers a
very precise test of the CKM unitarity condition
\bea
|V_{ud}|^2 + |V_{us}|^2 + |V_{ub}|^2 = 1.0001(10)~~~~~~\mbox{(90\% CL)}~,
\eea
which results in
\bea
\label{eq:real-CV}
{\rm Re}\left(\frac{\overline{C}_V + \overline{C}^\prime_V}{2}\right) = 1.000(1)~~~~~~\mbox{(90\% CL)}~,
\eea
in perfect agreement with the SM prediction.

%%%%%%%%%%%%%%%%%%%%%%%%%%%%%%%%%%%%%%%%%%%%%%%%%%%%%%%%%%%%%%%%%%%%%%%%%%%%%%%%%%%%%%%%%%%%
\subsection{Limits on real scalar and tensor couplings}
\label{sec:real-ST-nuclear}

\begin{figure}[!hbt]
\centerline{
\includegraphics[width=0.4\columnwidth]{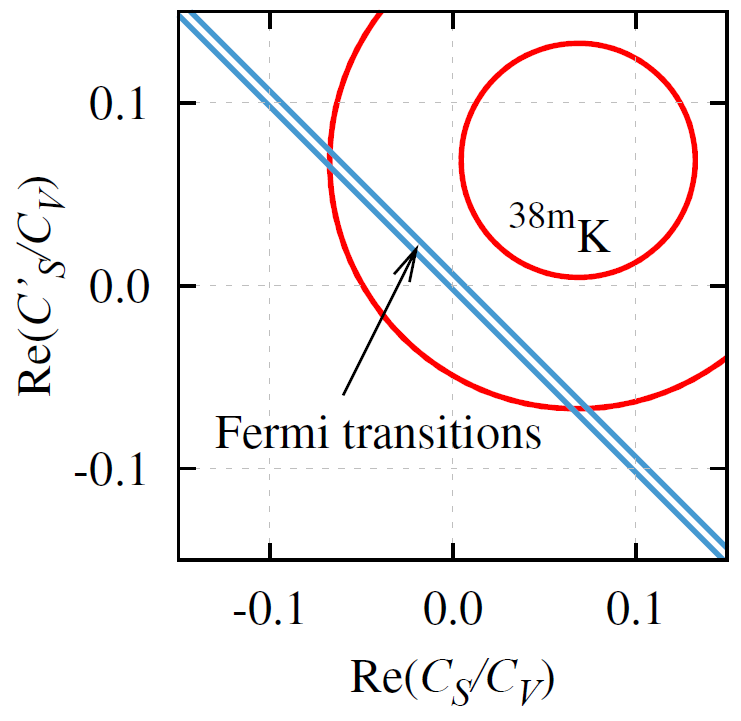}
}
\caption{
\label{fig:ReS-limNuclExp}
Constraints on real scalar couplings obtained from most precise
observables in nuclear $\beta$ decay. The straight lines are deduced
from the
Fierz interference term in super-allowed pure Fermi transitions
\cite{hardy09}. The circular bounds are deduced from the measurement of
the $\beta\nu$ angular correlation in $^{38m}$K decay \cite{gorelov05}.
The limits are calculated at the 90\% CL.
}
\end{figure}

The most stringent limits on scalar couplings obtained from nuclear
and neutron $\beta$ decays arise from the contribution of the Fierz
interference term to the $\mathcal{F}t$-values of
super-allowed pure Fermi transitions \cite{hardy09}.
The value of the Fierz term extracted from this set reads
\begin{equation}
b_F = -{\rm Re}\left( \frac{C_S + C_S^\prime}{C_V} \right) = -0.0022(43)~~~~~~\mbox{(90\% CL)}~,
\label{eq:bF}
\end{equation}
and the constraints are shown by the light blue lines in
Fig.~\ref{fig:ReS-limNuclExp}.
The error on this value is determined by the individual errors of the
$\mathcal{F}t$-values which are fitted to search for a possible deviation
from a constant.
The $Q_{EC}$ values in these transitions increase with the mass of the
parent nucleus so that transitions in lighter nuclei have a stronger weight
in the extraction of $b_F$ due to the $\langle m/E_e \rangle$ factor.
The contributions to the error on
$b_F$ given in Eq.(\ref{eq:bF}) are then due to the experimental data and
to theoretical corrections which have here an important effect.
The opportunities for improving the errors of the $\mathcal{F}t$-values
have recently been discussed in Ref. \cite{towner10} in connection with
the determination of $V_{ud}$ from pure Fermi transitions for
the unitarity test of the CKM matrix.

The other measured observable providing complementary
constraints to those resulting
from the ${\mathcal F}t$-values is the $\beta\nu$ angular correlation $a$.
The most precise result obtained so far was in the pure Fermi 
decay of $^{38m}$K. The experiment used the TRIUMF Neutral Atom Trap setup
\cite{gorelov05} which is a Magneto Optical Trap (MOT) system
composed of two traps.
The $\beta\nu$ correlation was determined from the shape of the time-of-flight
spectra of
recoil ions measured in coincidence relative to the $\beta$ particle.
Since $^{38m}$K is a positron emitter, the detection of positively
charged recoil ions relies on the double or multiple shake-off of
electrons following $\beta$ decay.
The statistical precision of the result is $3\times 10^{-3}$ and the systematic
error is comparable, arising from several instrumental sources \cite{gorelov05}.
A similar precision has been achieved with an indirect method, by measuring the
energy spectrum
shape of the delayed proton in the decay of $^{32}$Ar
\cite{adelberger99}.
The bounds obtained from the $^{38m}$K experiment are shown by the
circles in Fig.~\ref{fig:ReS-limNuclExp} and are extracted from
\bea
\label{eq:a-tilde}
\tilde{a} = \frac{a}{1+ b \langle m/E_e \rangle}~.
\eea
It is interesting to stress that the
abscissa, $x_C$, and ordinate, $y_C$, of the center of the circles
in Fig.~\ref{fig:ReS-limNuclExp}
are essentially given by $x_C \approx y_C \approx \langle m/E_e \rangle / 2$,
and that the mean radius of the circular band, $R_M$, is given by
$R_M \approx \langle m/E_e \rangle / \sqrt{2}$.
This indicates the importance of the value of the sensitivity factor
$\langle m/E_e \rangle$ of the Fierz term,
in defining the exclusion plot
and hence the interval of allowed values for the exotic couplings.

The value for a possible real scalar coupling obtained from the most recent
global analysis \cite{severijns06} including data from experiments in nuclear
and neutron decays translates into
\beq
\label{eq:CS-val-globFit}
{\rm Re}\left( \frac{C_S + C_S^\prime}{C_V} \right) = 0.0026(42)~~~~~~~\mbox{(90\% CL),}
%C_S/C_V = 0.0013(21)~~~~~~~\mbox{(90\% CL),}
\eeq
for a three parameters fit with left-handed couplings ($C_S = C^\prime_S$).
The fit also includes
data from ${\mathcal F}t$-values and the measurement of $\tilde{a}$ in $^{38m}$K
and $^{32}$Ar decays discussed above.
Although the compilation of ${\mathcal F}t$-values used in
Ref.~\cite{severijns06} was not the same than for the value quoted in
Eq.(\ref{eq:bF}), the result in Eq.~(\ref{eq:CS-val-globFit}) is clearly dominated
by the ${\mathcal F}t$-values.

\begin{figure}[!hbt]
\centerline{
\includegraphics[width=0.4\columnwidth]{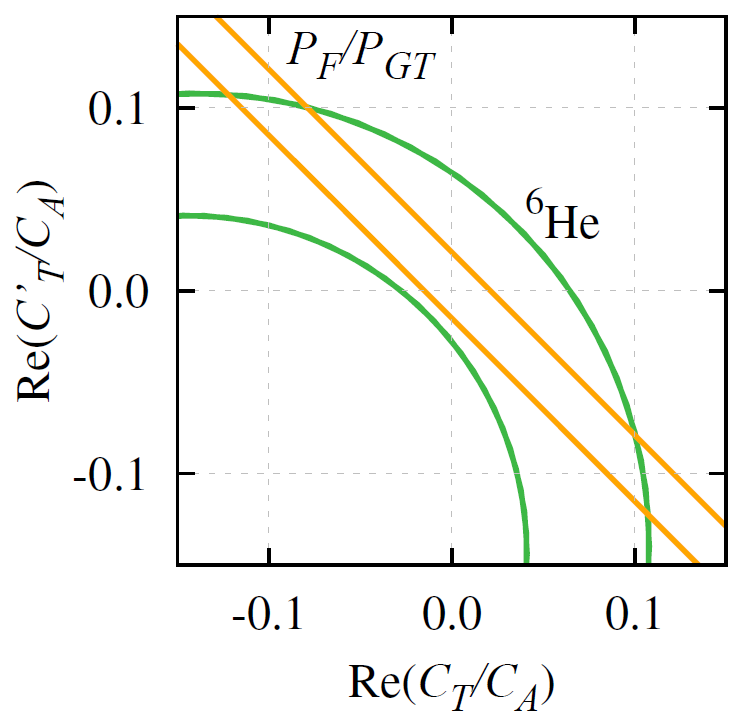}
}
\caption{
\label{fig:ReT-limNuclExp}
Constraints on real tensor couplings obtained from most precise single
observables in nuclear $\beta$ decay. The straight lines are deduced
from the
Fierz interference term contributing to the longitudinal
polarization of $\beta$ particles in Fermi and Gamow-Teller transitions
\cite{carnoy91}. The circular bounds are deduced from the measurement of
the $\beta\nu$ angular correlation in $^{6}$He decay \cite{johnson63}.
All limits are calculated at the 90\% CL.
}
\end{figure}

The most stringent limits on tensor couplings arising from single observables
is obtained from the ratio between polarizations
of $\beta$ particles emitted from pure Fermi and pure Gamow-Teller transitions
$P_{\rm F}/P_{\rm GT}$.
The longitudinal polarizations are governed by the correlation coefficients
$G$ of each decay, and the ratio provides constraints on both scalar and tensor
contributions. 
These experiments were motivated by the search for deviations from maximal
parity violation due to the presence of e.g. right-handed currents which would
couple through $V$ and $A$ interactions.
The polarization ratio $P_{\rm F}/P_{\rm GT}$ has been measured with high
precision by two groups
\cite{wichers87,carnoy90,carnoy91}.
The first experiment compared the longitudinal polarization of positrons
from $^{26m}$Al and $^{30}$P decays using Bhabha scattering in a magnetized foil
\cite{wichers87}.
The second group detected positrons from $^{10}$C and $^{14}$O \cite{carnoy90,carnoy91}
and used the polarimetry technique based on time-resolved spectroscopy of
hyperfine positronium decay.
Here again, it is the contribution of the Fierz term to $G$ that provides the
constraints on exotic couplings deduced from these experiments.
At the 90\% CL, the value deduced for the difference between the scalar and
tensor terms reads \cite{carnoy91}
\begin{equation}
{\rm Re} \left( \frac{C_S + C_S^\prime}{C_V} \right) -
{\rm Re} \left( \frac{C_T + C_T^\prime}{C_A} \right) = 0.003(18)~.
\label{eq:b-PFPGT}
\end{equation}
Since the scalar couplings are
more severely constrained by the ${\mathcal F}t$-values from pure Fermi transitions,
Eq.(\ref{eq:bF}),
these experiments provide constraints on the tensor contribution.
The limits are shown by the orange straight lines in
Fig.~\ref{fig:ReT-limNuclExp}.
It is remarkable that the most stringent limits on tensor couplings from nuclear
$\beta$ decay arise from two experiments that were performed in the 1980's,
even though that was not the main motivation of those experiments.

The $\beta\nu$ angular correlation provides also here complementary
constraints to those obtained from $P_F/P_{GT}$.
The most precise measurement carried out so far was in
the pure Gamow-Teller decay of $^6$He, fifty years ago \cite{johnson63}.
This experiment
measured the energy spectrum of the recoil ions in singles with an
electromagnetic spectrometer. The relative
precision achieved in this measurement was $10^{-2}$. The result
has been revisited by Gl\"uck to include order-$\alpha$ radiative
corrections and the effect of induced weak currents to the recoil
spectrum \cite{gluck98}.
The 90\% CL limits obtained from this measurement are shown by the
green circles on Fig.~\ref{fig:ReT-limNuclExp}.
It is seen that the (absolute) values of the coordinates $(x_C,y_C)$ of the
center of the circles and the mean radius, $R_M$, of the circular band
are here larger than in $^{38m}$K decay (Fig.~\ref{fig:ReS-limNuclExp}),
because the sensitivity factor $\langle m/E_e \rangle$ is
a factor 1.8 larger. Although the relative error on the measurement of
$\tilde{a}$ in $^6$He decay is a factor of about 2 larger than in $^{38m}$K,
the width of the circular band is smaller due its larger sensitivity.
As expected, the interval for the allowed values of
${\rm Re}(C_T + C^\prime_T)/C_A$ from the $^6$He experiment is dominated
by the contribution of the Fierz term to $\tilde{a}$.
Moreover, it appears that the sensitivity factors for the measurement of
$\tilde{a}$ in $^6$He and of $P_F/P_{GT}$ in $^{10}$C and $^{14}$O are the
same within 2\% \cite{severijns06}. Therefore the ratio between the bounds
on ${\rm Re}(C_T + C^\prime_T)/C_A$ is provided directly by the ratio
between the relative uncertainties of the experimental results.

The value for a possible real tensor coupling as deduced from the global analysis
of Ref.~\cite{severijns06}, for a three parameters fit with left-handed
couplings ($C_T = C^\prime_T$), is
\beq
\label{eq:CT-val-globFit}
{\rm Re}\left( \frac{C_T + C_T^\prime}{C_A} \right) = 0.007(11)~~~~~~~\mbox{(90\% CL)}~.
%C_T/C_A = 0.0036(54)~~~~~~~\mbox{(90\% CL)}~.
\eeq
The error resulting from this fit is somewhat smaller than the
one obtained in the measurements of $P_F/P_{GT}$ alone, Eq.~(\ref{eq:b-PFPGT}).
This is
attributed to the contribution of the Fierz term to the $\beta$ asymmetry parameter,
$\tilde{A}$, in neutron decay which has a significant impact on the global fit.
For the data used in Ref.~\cite{severijns06},
the sensitivity to the Fierz term through the factor $\langle m/E_e \rangle$
is about 2 times larger for the $\tilde{A}$ coefficient in neutron decay than
for $P_F/P_{GT}$ \cite{severijns06}.

Measurements of other correlations do not provide more stringent limits
compared to those presented above.
As already mentioned,
the coefficients which are linear in the real parts of exotic couplings and that
have been measured are $B$, $Q$ and $N$.

The coefficient $B$ has been measured in $^{37}$K
\cite{melconian07} and in neutron decay \cite{serebrov98,schumann07}. As
discussed in Sec.~\ref{sec:correlations}, the sensitivity of $\tilde{B}$ to exotic
couplings can be strongly suppressed due to the contribution of the Fierz term.
These experiments were however motivated by the search for right-handed currents
with vector and axial-vector couplings.

The same goal motivated measurements
of the longitudinal polarization of $\beta$ particles emitted from polarized nuclei,
which is dominated by the $Q$ coefficient and contains also a small contribution
of $N$ \cite{quin89}. The measurements were carried out in
$^{107}$In \cite{severijns93} and $^{12}$N
\cite{allet96,thomas01}. Since these are relative measurements, the contribution
of the Fierz term cancels and the remaining sensitivity to exotic left-handed
couplings is strongly suppressed.

Finally, the $N$ coefficient has been measured in neutron decay
\cite{kozela12} as a control parameter of a polarimeter dedicated to the
measurement of the triple correlation $R$. The achieved experimental error
combined with the small value
of $N$ under the conditions of that experiment do not provide significantly
improved constraints on scalar or tensor couplings.

Absolute measurements of the $\beta$ asymmetry parameter, $A$, 
have recently been performed in Gamow-Teller decays with the explicit purpose to
probe tensor couplings via the Fierz term \cite{wauters09,wauters10}.
The experiments used the low-temperature nuclear orientation technique in the
decays of 
$^{114}$In \cite{wauters09} and $^{60}$Co \cite{wauters10}. These transitions
have a relatively
low end point energies and hence large sensitivity to the Fierz term.
However, the allowed Gamow-Teller
matrix element is strongly reduced in $^{60}$Co decay. The SM value of the
asymmetry parameter has therefore a theoretical uncertainty which is of
the same order of magnitude than the experimental error \cite{wauters10} due to
the contribution of recoil corrections.
The total relative
uncertainties achieved were respectively 1.4\% and 2.0\% at $1\sigma$ for
$^{114}$In and $^{60}$Co and do not significantly improve the limits obtained
from $P_F/P_{GT}$.

In neutron decay, a global analysis of data has been performed
in order to assess the sensitivity to exotic couplings \cite{konrad11}.
It appears that, since the ratio $g_A/g_V$ has to be determined from the same
data, the constraints are less stringent than those obtained from the
global fit including nuclear
data \cite{severijns06}.
A similar conclusion was recently obtained in the analysis of $\tilde{a}$,
$\tilde{A}$, and $\tilde{B}$, in neutron decay \cite{holeczek13}.
It is to note that in the global fit of Ref.~\cite{severijns06},
the determination of the ratio $g_A/g_V$ results essentially from
the comparison between the ${\mathcal F}t$-value from pure Fermi transitions
and the neutron lifetime so that all correlation in neutron decay,
and in particular $A$ serve to constraint exotic couplings.

To summarize, we have shown that the most stringent limits obtained on the
real scalar and tensor couplings result solely from the contribution of
the Fierz term to the observables. This applies to the total decay rate
through the ${\mathcal F}t$-values, to the $\beta\nu$ angular correlation
coefficient $\tilde{a}$ in Fermi and Gamow-Teller transitions and to
the $\beta$ particle longitudinal
polarization $\tilde{G}$ which enters the ratio $P_F/P_{GT}$.

%%%%%%%%%%%%%%%%%%%%%%%%%%%%%%%%%%%%%%%%%%%%%%%%%%%%%%%%%%%%%%%%%%%%%%%%%%%%%%%%%%%%%%%%%%%%
\subsection{Limits on imaginary couplings}
\label{sec:imaginary_nuclear}

The presence of imaginary phases between the couplings is related to
the violation under time-reversal \cite{jackson57b}.
In nuclear and neutron decays, searches for time-reversal violation have
traditionally been focused on the $D$ and $R$ triple correlation coefficients
(Fig.~\ref{fig:correlations}).

Since the $\beta\nu$ angular correlation depends quadratically on the NP
couplings, it is also sensitive to possible imaginary parts in them.
Considering that measurements of the ${\mathcal F}t$-values in Fermi
transitions and of the polarization ratio $P_{\rm F}/P_{\rm GT}$, provide
stringent constraints on the Fierz terms associated respectively with
scalar,
and tensor couplings, Eqs.~(\ref{eq:bF}) and (\ref{eq:b-PFPGT}), it is
possible to extract constraints on the imaginary parts from the
$\beta\nu$ correlation measurements, by neglecting here the contributions of
the Fierz terms.

\begin{figure}[!htb]
\centerline{
\includegraphics[width=0.4\columnwidth]{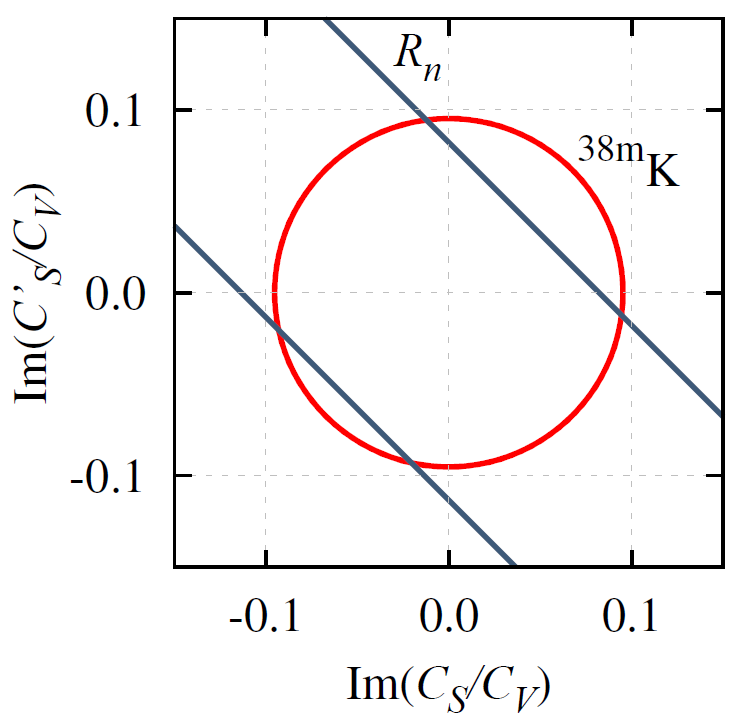}
}
\caption{
\label{fig:ImS-limNuclExp}
Constraints on imaginary scalar couplings obtained from most precise
single experiments in nuclear and neutron decays. The straight lines
are deduced from the triple correlation measurement in neutron decay
\cite{kozela12}. The circular bound is deduced from the measurement of
the $\beta\nu$ angular correlation in $^{38m}$K decay \cite{gorelov05}.
The limits are calculated at the 90\% CL.
}
\end{figure}

Figure \ref{fig:ImS-limNuclExp} shows the constraints on imaginary
scalar couplings extracted from the $\beta\nu$ angular correlation
in $^{38m}$K decay \cite{gorelov05} and from the measurement of the
triple correlation coefficient, $R$, in neutron decay \cite{kozela12}.
The determination of this coefficient requires the measurement of the
transverse polarization of $\beta$ particles emitted perpendicular to
the neutron spin.
The experiment was performed using a polarized cold neutron beam
and the transverse electron polarization was analyzed
from the asymmetry in Mott scattering at backward angles
using a thin lead foil.
The total absolute error reached in this measurement is $\delta R_n = 1.3$\%,
generating the following 90\% CL constraint
\begin{equation}
\label{eq:ImCS}
{\rm Im} \left( \frac{C_S + C^\prime_S}{C_V} \right)
 -1.5~ {\rm Im} \left( \frac{C_T + C^\prime_T}{C_A} \right) = -(1.8\pm 5.9) \times 10^{-2}~.
%{\rm Im}(C_S + C^\prime_S)/C_V = (0.xx\pm 0.xx) \times 10^{-2}~.
\end{equation}
Using the stronger constraint on the tensor contribution obtained from the
measurement of $R$ in $^8$Li decay (see below), we obtain the limit on the imaginary
part of the scalar interaction shown in Fig.~\ref{fig:ImS-limNuclExp}.

Notice that the quadratic dependence of the $\beta\nu$ angular correlation provides here
a competitive constraint.
The region allowed by the $^{38m}$K result (disk inside the red circle)
has been somewhat reduced by the constraint obtained from the new measurement
of the triple correlation in neutron decay.

\begin{figure}[!hbt]
\centerline{
\includegraphics[width=0.4\columnwidth]{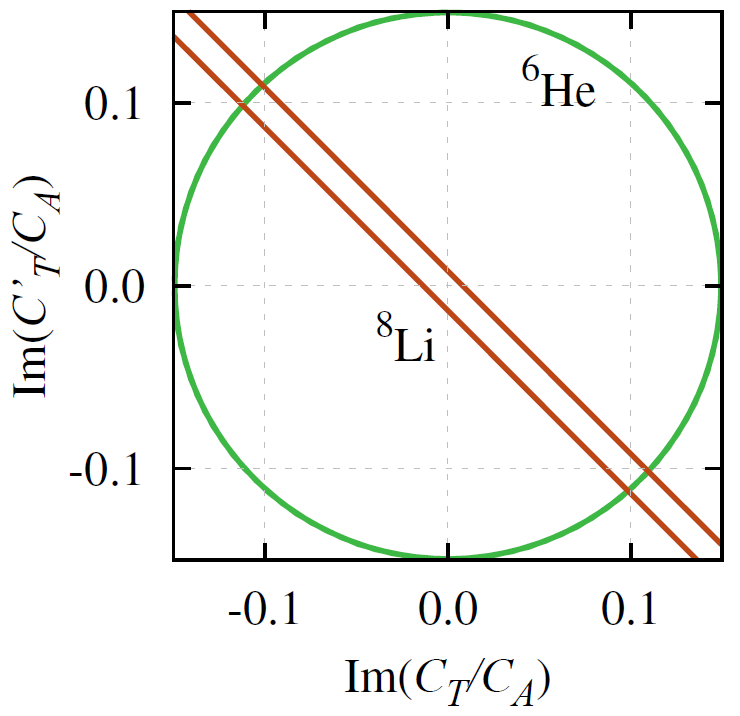}
}
\caption{
\label{fig:ImT-limNuclExp}
Constraints on imaginary tensor couplings obtained from most precise
single experiments in nuclear $\beta$ decay. The straight lines
are deduced from the triple correlation measurement in $^8$Li decay
\cite{huber03}. The circular bound is deduced from the measurement of
the $\beta\nu$ angular correlation in $^{6}$He decay \cite{johnson63}.
The limits are calculated at the 90\% CL.
}
\end{figure}

The constraints on the imaginary tensor couplings are shown in
Fig.~\ref{fig:ImT-limNuclExp}. They are extracted from the $\beta\nu$
angular correlation
in $^{6}$He decay \cite{johnson63} and from the measurement of the
triple correlation coefficient $R$ in $^8$Li decay \cite{huber03}.
This measurement used polarized $^8$Li nuclei produced by polarization
transfer reactions from a vector-polarized
deuteron beam on a $^7$Li target. The target was cooled
close to liquid helium temperatures in order to achieve long
relaxation times. The
transverse polarization of the decay electrons was also deduced here
from the Mott scattering asymmetry at backward angles
using a lead foil as analyzer. The absolute precision reached in this
measurement was $\delta R(^8{\rm Li}) = 2.2\times 10^{-3}$ and the result has been
corrected for the
effects of final state interactions, $R_{FSI}=0.7(1) \times 10^{-3}$.
The 90\% CL value of the imaginary part of tensor couplings obtained
from this measurement reads
\begin{equation}
\label{eq:ImCT}
{\rm Im} \left( \frac{C_T + C^\prime_T}{C_A} \right) = (0.27\pm 1.08) \times 10^{-2}~.
\end{equation}

The situation for the constraints on tensor couplings, Fig.~\ref{fig:ImT-limNuclExp},
is quite different from the scalar couplings exclusion plot. The strongest
constraints are
those arising from the measurement of $R$, which has an uncertainty
a factor of 6 smaller than in neutron decay.

The measurement of the $D$ triple correlation coefficient requires the use
of mixed transitions and the determination of the neutrino
momentum through the observation of the recoil.
In nuclear $\beta$ decay, the $D$ coefficient has been measured in
$^{19}$Ne decay. The combined result from all runs provided the value
$D(^{19}{\rm Ne}) = (1\pm6)\times 10^{-4}$ \cite{calaprice85}.
The $D$ coefficient has also been measured by two groups in neutron
decay \cite{soldner04,Chupp:2012ta}. The most recent and precise 
result is  $D_n = (-0.9\pm 2.1)\times 10^{-4}$
\cite{Chupp:2012ta} and provides the following 90\% CL bounds for a possible
phase between the vector and axial couplings
\begin{equation}
\label{eq:ImCVCA}
{\rm Im}(C_V C^*_A) = (-2.1\pm 8.0) \times 10^{-4}~.
\end{equation}

This completes the review of the current status of the six couplings left
in the approximation introduced in Sec.~\ref{sec:quark-level-Leff},
that affect linearly the observables.

%%%%%%%%%%%%%%%%%%%%%%%%%%%%%%%%%%%%%%%%%%%%%%%%%%%%%%%%%%%%%%%%%%%%%%%%%%%%%%%%%%%%%%%%%%%%
\section{Comparison with other low- and high-energy experiments}
\label{sec:lowE-LHC}

\subsection{CP-conserving vector/axial couplings}
As explained in Sec.~\ref{sec:quark-level-Leff}, the linear effects
produced by CP-conserving vector and axial-vector interactions, which are
represented by $\epsilon_{L,R}$ and $\tilde{\epsilon}_{L,R}$
can be re-expressed as an overall shift
in the effective Lagrangian and an unobservable shift in the
axial-vector form factor $g_A$. The only observable effect is then
a NP contribution to $V_{ud}$ that can be probed through the test
of the unitarity condition of the CKM matrix.

The value obtained in Eq.(\ref{eq:real-CV}) translates into the
following strong bound
\bea
\label{eq:LpR}
|\mbox{Re} (\eL + \eR)| < 0.5\times 10^{-3}~~~~~~\mbox{(90\% CL)}~,
\eea 
where once again we have neglected ${\cal O}(\epsilon_i, \tilde{\epsilon}_i)^2$ contributions. In the high-energy effective theory, this bound corresponds to an effective NP scale of 11 TeV that represents a more stringent bound than those from LEP and LHC analysis on the same effective interactions \cite{Cirigliano:2009wk,Cirigliano:2012ab}.

Given a specific NP model this stringent bound translates in severe constraints on the masses and couplings of the new particles, as explicitly shown in various (mostly supersymmetric) extensions of the SM during the last decades~\cite{Barbieri:1985ff,Marciano:1987ja,Hagiwara:1995fx,Kurylov:2001zx,Marciano:2007zz,Bauman:2012fx}.

%%%%%%%%%%%%%%%%%%%%%%%%%%%%%%%%%%%%%%%%%%%%%%%%%%%%%%%%%%%%%%%%%%%%%%%%%%%%%%%%%%%%%%%%%%%%
\subsection{CP-conserving scalar and tensor couplings}
\label{sec:CP-even-ST}
The rest of non-standard interactions can be probed by measuring normalized angular and energy distributions, as described in Sec.~\ref{sec:nuclear}. %The results presented there illustrate our discussion in Sec.~\ref{sec:quark-level-Leff} about the NP linear contributions, since we can see that the bounds on $\mbox{Re}(\epsilon_{S,T}) \sim C_{S,T} + C'_{S,T} $ are significantly stronger than for the rest of the couplings.

%An exhaustive analysis of the NP bounds obtained from a long list of nuclear and neutron beta decay measurements was performed in Ref.~\cite{severijns06}. Bounds on all six CP-conserving coefficients Re($\epsilon_{S,T}$) and Re($\tilde{\epsilon}_{S,T,L,R}$) were obtained. As expected from our discussion in Sec.~\ref{sec:quark-level-Leff} about the NP linear contributions, only the bounds on $\mbox{Re}(\epsilon_{S,T})$ are below the percent level. Now we take a closer look at the current and future experiments probing these two couplings.

\begin{figure}[!hbt]
\centerline{
\includegraphics[width=0.5\columnwidth]{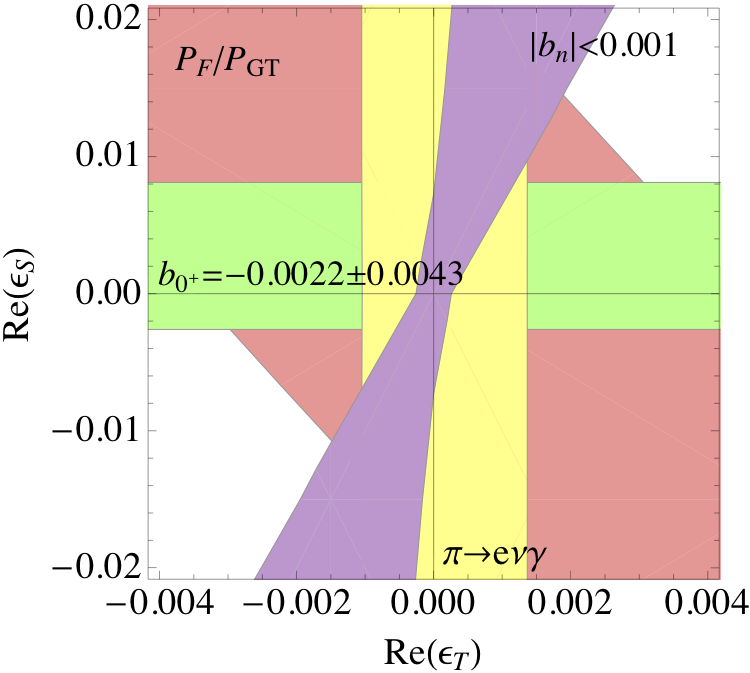}
}
\caption{\label{fig:low-E-constraints}
90\% CL limits on the scalar and tensor NP couplings Re($\epsilon_{S,T})$ from super-allowed nuclear decays \cite{hardy09} (green), radiative pion decay \cite{Bychkov:2008ws} (yellow) and measurements of the ratio $P_{\rm F}/P_{\rm GT}$~\cite{carnoy91,wichers87} (red), along with the expected bound from future measurements of the Fierz term $b$ in neutron decay (purple).
}
\end{figure}

For scalar interactions, the most stringent bound comes from the determination of the Fierz
term in super-allowed Fermi decays given in Eq.\eqref{eq:bF}. This translates into
\beq
 - 1.0 \times 10^{-3}  <   g_S \, \mbox{Re}(\eS)  <  3.2 \times 10^{-3}  
~~~~~~~\mbox{(90\% CL)}~,
\label{eq:bFbound}
\eeq
represented in Fig.~\ref{fig:low-E-constraints} with a horizontal green band. 

In the case of tensor interactions, we saw in Sec.~\ref{sec:real-ST-nuclear} that the best bound from nuclear $\beta$
decays arises from measurements of the ratio $P_{\rm F}/P_{\rm GT}$ of longitudinal polarization of positrons emitted in the decay of pure Fermi and pure Gamow-Teller transitions~\cite{carnoy91,wichers87}. These experiments are sensitive to the difference between Fierz terms and generate the following 90\% CL bound
\bea
- 2.6 \times 10^{-3}  \ < \    \frac{g_S}{3} \, \mbox{Re}(\eS)   +    g_T \, \mbox{Re}(\eT) \  <  3.2 \times 10^{-3} ~.
\eea
This is represented in Fig.~\ref{fig:low-E-constraints} by the wide red diagonal band.
The contribution of other observables in nuclear and neutron decays that can
possible improve this limit require a global fit as
presented in Ref.~\cite{severijns06}.

More stringent bounds on tensor interactions can be obtained from the analysis of the Dalitz plot of the radiative pion decay $\pi^+  \to  e^+ \nu_e \gamma$ done by the PIBETA collaboration~\cite{Bychkov:2008ws}\footnote{For the associated hadronic form factor we use $f_T = 0.24(4)$, obtained in Ref.~\cite{Mateu:2007tr}, using a large-$N_c$-inspired resonance-saturation model.%$f_T = 0.24(4)$ at the renormalization scale $\mu=1$~GeV, with parametric uncertainty induced by the uncertainty in the quark condensate. 
}
\beq
- 1.1  \times 10^{-3}  \ < \ \mbox{Re}(\eT) \  <  \  1.4  \times 10^{-3}~~~~~~~\mbox{(90\% CL)}~,
\label{eq:tradpi}
\eeq
represented in Fig.~\ref{fig:low-E-constraints} by the vertical yellow band. 

Another promising process to probe both scalar and tensor interactions is the measurement of the Fierz term in neutron decay~\cite{Bhattacharya:2011qm}. The purple area in Fig.~\ref{fig:low-E-constraints} shows the impact of future determinations of this parameter with a sensitivity at the level of $|\delta b| \leq 10^{-3}$. It will offer the most stringent bound on tensor interactions, after taking into account the strong bounds existing on $b_{\rm F}$, Eq.~(\ref{eq:bF}).

Finally it is worth mentioning that, although not shown in Fig.~\ref{fig:low-E-constraints}, the ratio  $\rm{R_\pi = \Gamma(\pi\to e\nu) /\Gamma(\pi\to \mu\nu)}$ is also a very powerful probe of scalar and tensor interactions since they generate radiatively a pseudo-scalar interaction \cite{Campbell:2003ir}. More details can be found in Refs.~\cite{Bhattacharya:2011qm,Cirigliano:2012ab}.

%%%%%%%%%%%%%%%%%%%%%%%%%%%%%%%%%%%%%%%%%%%%%%%%%%%%%%%%%%%%%%%%%%%%%%%%%%%%%%%%%%%%%%%%%%%%
\subsection{CP-violating interactions}

The three CP-violating phases with a linear effect on the nuclear $\beta$ decay observables are represented in the quark-level effective Lagrangian by the imaginary parts of the coefficients $\epsilon_{R,S,T}$. They parameterize the relative phase between the purely vector interaction in the hadronic bilinear and axial-vector, scalar and tensor interactions, respectively.

The 90\% CL bounds on the imaginary parts of scalar and tensor interactions obtained from measurements of
the $R$ parameter in neutron and in $^8$Li decays were given in Eqs.\eqref{eq:ImCS} and \eqref{eq:ImCT}. They can be trivially re-expressed as
\bea
g_S~\mbox{Im}(\eS) + 4.7~g_T~\mbox{Im}(\eT) &=& -(0.9 \pm 3.0)\times 10^{-2}
\label{eq:ImeS} \\
g_T~\mbox{Im}(\eT) &=& -(0.4 \pm 1.7)\times 10^{-3}~.
\label{eq:ImeT}
\eea

\begin{figure}[!hbt]
\centerline{
\includegraphics[width=0.5\columnwidth]{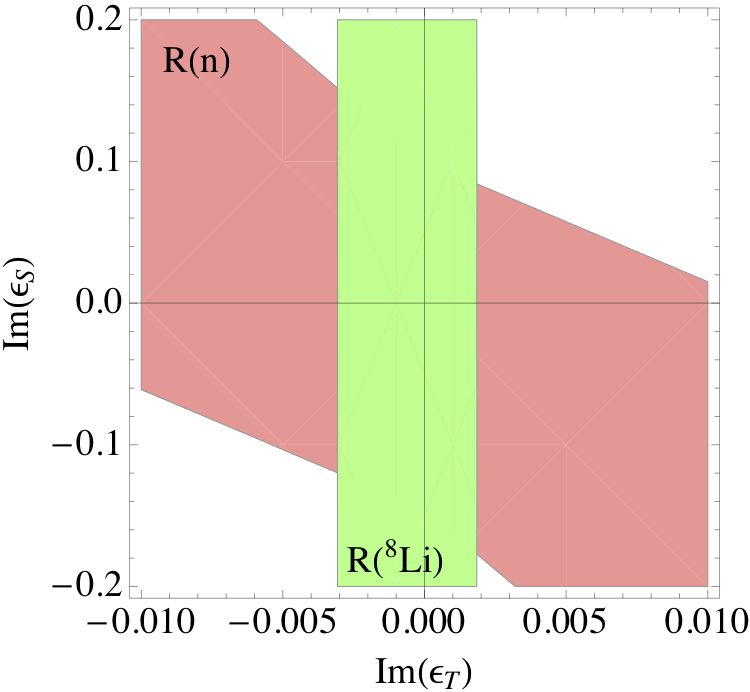}
}
\caption{\label{fig:low-E-constraints-complex}
90\% CL limits on the CP-violating scalar and tensor NP couplings Im($\epsilon_{S,T})$ from measurements of the triple correlation R in neutron decay \cite{kozela12} (diagonal maroon band) and $^8$Li \cite{huber03} (vertical green band). Notice that both $\eS$ and $\eT$ scales are different than in Fig.~\ref{fig:low-E-constraints}.}
\end{figure}

Using the recent lattice QCD determination of the form factors $g_{S,T}$~\cite{Bhattacharya:2011qm} we obtain the bounds shown in Fig~\ref{fig:low-E-constraints-complex}.

Likewise, the bound on the relative phase between $C_V$ and $C_A$ given in Eq.\eqref{eq:ImCVCA} from the measurement of the $D$ correlation coefficient in neutron decay \cite{Chupp:2012ta} can be casted in the quark-level language as
\bea
\mbox{Im}(\eR) &=& -(1.1 \pm 4.0)\times 10^{-4}~~~~~~~\mbox{(90\% CL)}~.
\label{eq:ImeR}
\eea
 
%
%The measurements of the $R$ coefficient in pure Gamow-Teller transitions give strong bounds on the CP-violating tensor coupling Im($\eT$). On the other hand, bounds on the scalar coupling Im($\eS$) are much weaker, simply because we do not have such a precise measurement of the $R$ coefficient in a mixed transition with large Fermi contribution.
%
It is worth mentioning that additional T-odd correlations with potential NP sensitivity can be constructed
in the radiative $\beta$ decay of nuclei and neutron, as shown in Refs.~\cite{Gardner:2013aiw,Gardner:2012rp}.
%Refs.~\cite{Gardner:2013aiw,Gardner:2012rp} studied an additional T-odd correlation that can be constructed in the radiative beta decay of nuclei and neutron, computing the FSI effects in such a way that future measurements could be used to search for CP-violating NP.

Like for the CP-conserving coefficients, the ratio  $\rm{R_\pi = \Gamma(\pi\to e\nu) /\Gamma(\pi\to \mu\nu)}$ offers strong constraints on $\mbox{Im}(\epsilon_{S,T})$ since they generate radiatively a non-zero $\mbox{Im}(\epsilon_P)$ \cite{Cirigliano2013}.

Using the high-energy effective Lagrangian of Eq.\eqref{eq:BW}, it is possible to show that the same $SU(2)_L\times U(1)_Y$ invariant effective operators that generate at low-energy the coefficients $\epsilon_{R,S,T}$ also generate contributions to different EDMs~\cite{Ng:2011ui}, what generates much stronger bounds than those given in Eqs.\eqref{eq:ImeS}-\eqref{eq:ImeR}. %obtained from nuclear and neutron decay studies. 
In fact, the indirect limit on the $D$ coefficient from the current neutron EDM bound is of the order $10^{-7}$~\cite{Ng:2011ui}, whereas for the $R$ coefficient the current Thallium EDM bound implies an indirect bound at the level of $10^{-8}$~\cite{Tulin}. These bounds from EDMs could be avoided assuming an almost complete cancellation with other effective operators contributing to the EDMs.
Although such a scenario is very unnatural from a purely EFT point of view, in a specific NP model the different Wilson coefficients are related to the more fundamental coupling constants and masses and such a cancellation could occur in a less unnatural way. In this sense, direct bounds from $\beta$ decays complement EDM experiments in the search of new sources of CP-violation.
Moreover, this comparison with EDM relies on the use of the high-energy effective Lagrangian of Eq.\eqref{eq:BW}, that in turn relies on some assumptions about the structure of the underlying NP.

We can see that the situation is very different from the CP-conserving coefficients,
where direct limits from $\beta$ decays are very competitive and for some interactions
they actually offer the best bound.  

%
%\begin{itemize}
%\item Relative phase between V and A = Im($\epsilon_R)$. The form factor does not have a phase! Discussion of the best bound (EmiT in neutron?) and best nuclear bound (D in $^6Ne$?).
%\item Relative phase between V and T = Im($\eT)$. Discussion of the best bound (R in GT?).
%\item Relative phase between V and S = Im($\eS)$. Discussion of the best bound (R in F?). It is so weak that one actually gets a better limit from quadratic contributions to CP-conserving observables like $a_{\rm GT}$. Discuss it.
%\item Work by Tulin-Ng about the relation with EDMs. It is possible to derive very strong bounds, several orders of magnitudes stronger, unless some unnatural cancellations take place between operators in EDMs.
%\end{itemize}

%%%%%%%%%%%%%%%%%%%%%%%%%%%%%%%%%%%%%%%%%%%%%%%%%%%%%%%%%%%%%%%%%%%%%%%%%%%%%%%%%%%%%%%%%%%%
\subsection{Limits from the LHC}
If the new particles are too heavy to be produced on-shell at the LHC we can %nicely  (Deja que eso lo digan otros...)
connect collider searches with low-energy experiments in an elegant % Bueno... este lo dejo
model-independent way using the high-energy effective Lagrangian of Eq.~\eqref{eq:BW} to analyze collider data. 
The natural channel to study at the LHC is the search for electrons and missing transverse energy (MET),
$pp\to e+ {\rm MET}+X$, since the underlying partonic process is the same as in $\beta$ decay ($\bar{u}d\to e\bar{\nu}$) and so we expect it to be sensitive to the same kind of NP.

Using the matching conditions between Wilson coefficients of the low- and high-energy effective Lagrangians~\cite{Bhattacharya:2011qm,Cirigliano:2012ab} it is possible to express collider observables in terms of the coefficients of the low-energy effective theory, $\epsilon_i$ and $\tilde{\epsilon}_i$. In particular the cross-section $\sigma(pp\to e+{\rm MET}+X)$ with transverse mass higher than $\overline{m}_T$ takes the following form\footnote{Notice that high-energy searches probe separately the vertex correction $\eLv$ and contact $\eLc$ contributions to the coupling $\eL$, defined in Ref.~\cite{Cirigliano:2012ab}.}:
%\bea
%\label{eq:sigmamt}
%\sigma(m_T \!\!>\! \overline{m}_T) &=&    
%\sigma_W \Big[ 1  + | \eR|^2 \Big] +~ \sigma_R  \Big[ |\teR|^2 \!+ |\eLc|^2 \Big]
%\nonumber\\
%&&\hspace{-1.5cm}
%+ \sigma_S \Big[ |\eS|^2  \!+  |\teS|^2  \!+ |\eP|^2 \!+  |\teP|^2   \Big] +  \sigma_T  \Big[ |\eT|^2  \!+  |\teT|^2     \Big]~,\nonumber 
%\eea 
\bea
\label{eq:sigmamt}
\sigma(m_T \!\!>\! \overline{m}_T) &=&
\sigma_W \Big[ \Big| 1 +   \eLv \Big|^2+  |\teL|^2  + | \eR|^2 \Big] 
\\
&&\hspace{-1.5cm}
- 2 \, \sigma_{WL}\, \mbox{Re} \left( \eLc  +  \eLc \eLv^* \right) +~ \sigma_R  \Big[ |\teR|^2 \!+ |\eLc|^2 \Big]
\nonumber\\
&&\hspace{-1.5cm}
+ \sigma_S \Big[ |\eS|^2  \!+  |\teS|^2  \!+ |\eP|^2 \!+  |\teP|^2   \Big] +  \sigma_T  \Big[ |\eT|^2  \!+  |\teT|^2     \Big]~,\nonumber 
\eea  
%Here $\sigma_W (\bar{m}_T)$ is the SM contribution, so we can see that the LHC is not very sensitive to $\eLv$, $\teL$ and $\eR$, whereas $\sigma_{WL, R, S,T}(\bar{m}_T)$ are several orders of magnitudes larger than the SM contribution, which explicit form can be found in Ref.~\cite{REF}. 
where $\sigma_W (\overline{m}_T)$ represents the SM contribution and $\sigma_{WL, R, S,T}(\overline{m}_T)$ are new functions, which explicit form can be found in Ref.~\cite{Cirigliano:2012ab}. The crucial feature is that they are several orders of magnitudes larger than the SM contribution, what compensates for the smallness of the NP couplings and makes possible to put significant bounds on them from these searches. On the other hand, the lack of such an enhancement makes this search not very sensitive to $\eLv$, $\teL$ and $\eR$.

\begin{figure}[!hbt]
\centerline{
\includegraphics[width=0.5\columnwidth]{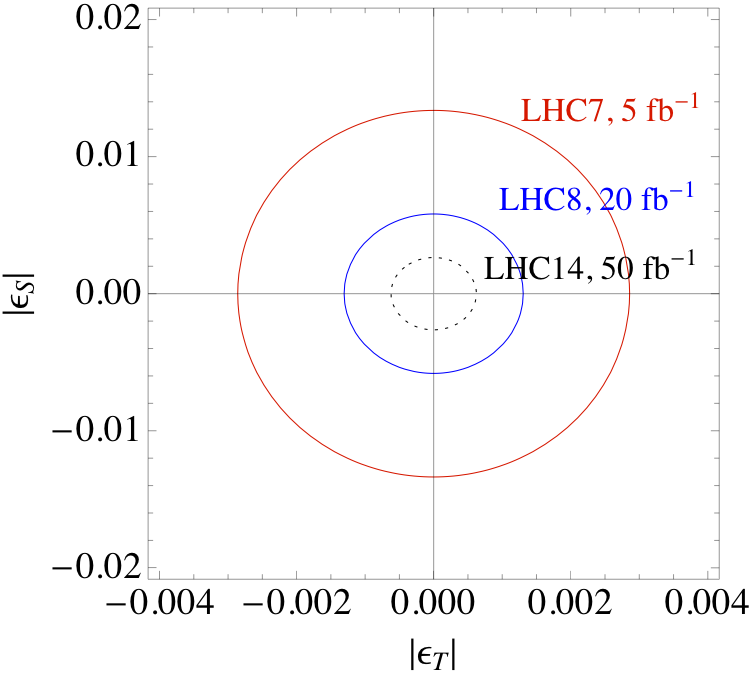}
}
\caption{
\label{fig:LHC-constraints}
The red (blue) solid line gives the 90\% C.L. limits on the scalar and tensor NP couplings $|\epsilon_{S,T}|$ obtained with 5 fb$^{-1}$ (20 fb$^{-1}$) of data recorded at $\sqrt{s}=$ 7 TeV (8 TeV) by the CMS collaboration in the $pp\to e+{\rm MET}+X$ channel \cite{ENcms5fb,ENcms20fb}. The dotted line gives an estimated future bound obtained with higher luminosity and energy.
}
\end{figure}

The most recent search for electrons and missing transverse energy in the final state was done by the CMS collaboration using 20 fb$^{-1}$ of data recorded at $\sqrt{s}=$ 8 TeV~\cite{ENcms20fb}. In this analysis they found one single event with a transverse mass above $1.5$ TeV, to be compared with the SM expectation of $1.99\pm 0.27$ events. This absence of an excess of high-$m_T$ events in this channel can be translated into bounds on the different NP couplings using Eq.~\eqref{eq:sigmamt}, as shown in Fig.~\ref{fig:LHC-constraints} for scalar and tensor couplings.%, where the previous bound obtained in Ref.~\cite{ENcms5fb} with 5 fb$^{-1}$ of data recorded at $\sqrt{s}=$ 7 TeV by the CMS Coll. 

%From Eq.~\eqref{eq:sigmamt} it is clear that the bounds for $\epsilon_{S,P}$ and $\tilde{\epsilon}_{S,P}$ are the same, and likewise for $\eT$ and $\teT$. Therefore if we assume that only one operator is present we have the following current 90\% C.L. bounds
Assuming that only one operator at a time is present we obtain the following 90\% CL bounds
\bea
%\label{eq:}
&&|\epsilon_{S,P}|, |\tilde{\epsilon}_{S,P}| < 5.8\times 10^{-3} ~,\\
&&|\eT|, |\teT| < 1.3\times 10^{-3}~,\\
&&|\teR|,|\mbox{Im}~\eLc| < 2.2\times 10^{-3}~,\\
&&\mbox{Re}~\eLc \in (-1.1,4.5)\times 10^{-3}~.
\eea

The bounds presented in Figs.~\ref{fig:low-E-constraints} and \ref{fig:LHC-constraints} show an interesting competition between low- and high-energy searches looking for
new CP-conserving scalar and tensor interactions involving LH neutrinos. On the other hand, interactions involving RH neutrinos are more strongly constrained by the LHC, as shown in Table~\ref{tab:summaryR}.%~\cite{Cirigliano:2012ab}.

In the case of non-standard (axial-)vector interactions with LH neutrinos, the combination Re($\eL+\eR$) is strongly constrained by CKM unitarity tests, as shown in Eq.~\eqref{eq:LpR}, but the orthogonal combination Re($\eL-\eR$) cannot be probed by $\beta$ decay experiments. 
 Although current LHC searches cannot improve the CKM unitarity limit, they are sensitive to the contact interaction part of $\eL$ providing in this way complementary information. % \cite{Cirigliano:2012ab}. 
 For (axial-)vector interactions with RH neutrinos, we see from Table~\ref{tab:summaryR} that the LHC dominates the search in the case of $\teR$, whereas the bounds obtained on $\teL$ from nuclear $\beta$ decays are the most competitive, even if they are above the per-cent level.

Finally, for the pseudoscalar couplings $\eP$ and $\teP$ the ratio $\rm{R_\pi = \Gamma(\pi\to e\nu) /\Gamma(\pi\to \mu\nu)}$ represents by far the best probe, providing bounds of order $10^{-4}$~\cite{Cirigliano:2012ab}.

All in all, we see that only the combination of both low- and high-energy
searches can give us a complete picture of non-standard charged current interactions.

\begin{table}
\centering
\caption{
\label{tab:summaryR}
Summary of 90\% CL bounds (in units of  $10^{-2}$) on the non-standard couplings $\tilde{\epsilon}_i$ from low- and high-energy searches.%~\cite{Bhattacharya:2011qm}.
}
\begin{tabular}[]{|c|c|c|c|c|c|}  
\hline\hline
%				& $\epsilon_L + \epsilon_R$	&	$\epsilon_L - \epsilon_R$			&	$\epsilon_S $	&	$\epsilon_P$			&	$\epsilon_T$      \\              
				&	$ |  \teL| $				&	$| \teR| $           				&	$| \teP|$			&	$| \teS|$			&	$|\teT|$               \\
Low energy \cite{severijns06,Cirigliano:2012ab}	&		6					&		6						&	0.03				&		14			&	3.0				\\
%LHC $(e \nu)$		&		-					&	0.5							&	1.3				&		1.3			&	0.3				\\%LHC7-5fb^-1
LHC (this work)			&		-					&	0.2							&	0.6				&		0.6			&	0.1				\\%LHC8-20fb^-1
\hline\hline
  \end{tabular}
\end{table}
%%% ... and for for free-style tabular material that won't fit easily in the

Needless to say, this interplay becomes much more interesting if a non-zero result is obtained for one of the Wilson coefficients, either in the low-energy experiments or in collider searches. This was explained in Ref.~\cite{Bhattacharya:2011qm}, where it was shown that a hypothetical scalar resonance found at the LHC in the $pp\to e^\pm+{\rm MET}+X$ channel would imply a lower bound in the value of $|\epsilon_S|$ that should then be confirmed in nuclear and neutron decay experiments. 

%%%%%%%%%%%%%%%%%%%%%%%%%%%%%%%%%%%%%%%%%%%%%%%%%%%%%%%%%%%%%%%%%%%%%%%%%%%%%%%%%%%%%%%%%%%%
\section{Experimental activities}

Detailed accounts of new results and ongoing activities
have been provided in several recent reviews
\cite{severijns06,severijns11,severijns13,abele08,nico09,dubbers11}.
In this section we focus on current experiments or projects
aiming at improving present limits on exotic couplings for both the
real and imaginary parts.
The purpose is to review the anticipated precision goals of such
efforts in order to confront them with the current most precise limits
and with the estimated reach at the LHC.

%%%%%%%%%%%%%%%%%%%%%%%%%%%%%%%%%%%%%%%%%%%%%%%%%%%%%%%%%%%%%%%%%%%%%%%%%%%%%%%%%%%%%%%%%%%%
\subsection{The $\beta\nu$ angular correlation, $a$}

As already indicated,
the information about the angular correlation between the $\beta$
particle and
the neutrino is contained in the momentum spectrum of the recoiling
daughter nucleus.
This correlation can be obtained by detecting the recoiling nuclei either
in singles or in coincidence with the $\beta$ particles.

There are currently two experiments aiming at improving the precision on
scalar couplings in Fermi transitions.

The first uses the Weak Interaction Trap for CHarged particles
spectrometer (WITCH) installed at ISOLDE-CERN \cite{beck03}.
The setup includes two Penning traps located in the same magnetic field. The
first trap serves for cleaning and preparation
of the ion cloud and the second to store the ions during their decay.
In the WITCH spectrometer,
the energy spectrum of the recoil ions is measured in
singles using a magnetic spectrometer which contains several
electrostatic retardation electrodes.
The operation of this type of spectrometer is complicated by the possible presence
of local Penning traps, where ions or electrons can be confined. This is
caused by the electromagnetic field
configuration inside the system and by secondary ionization processes. Such effects
require the operation of the spectrometer under ultra-high vacuum conditions.
The candidate nucleus considered for this experiment is $^{35}$Ar which is
also a positron
emitter. The detection of positively charged recoil ions
from singly charged trapped ions relies on the single or multiple
shake-off process of the bound electrons.
A first measurement of a recoil spectrum from stored $^{124}$In$^+$ ions has been
reported \cite{beck11}. The precision goal of this experiment is to reach
the level of $\delta a \approx 5\times 10^{-3}$ \cite{beck03}\footnote{The projections
of the future experimental precision goals do not always specify the statistical
CL. Unless explicitly stated otherwise, it is assumed that the quoted experimental
errors and the projected precision goals are at 1$\sigma$.}.
When achieved,
this will slightly improve the limit on scalar interactions obtained from $^{38m}$K,
Fig.~\ref{fig:ReS-limNuclExp}, and $^{32}$Ar decays.

In order to calculate the level of precision required in a measurement of $\tilde{a}_F$
from a pure Fermi transition, such as to compete with the direct
extraction of the Fierz term from the $\mathcal{F}t$-values in super-allowed pure
Fermi transitions, Eq.~\eqref{eq:bF}, we
refer once again to the only linear sensitivity of $\tilde{a}$ to scalar interactions
through the contribution of the Fierz term, Eq.(\ref{eq:a-tilde}). This leads to
\bea
\label{eq:a-exp}
\left| \frac{\delta \tilde{a}_F}{\tilde{a}_F}\right| \approx |\delta b_F| \big\langle \frac{m}{E} \big\rangle~.
\eea
Taking $\delta b_F$ at 1$\sigma$ from Eq.~\eqref{eq:bF} and
assuming $\langle m/E \rangle \sim 0.2$ as an estimate, Eq.~(\ref{eq:a-exp}) gives
$|\delta \tilde{a}_F / \tilde{a}_F| \sim 0.5\times 10^{-3}$.
This is an order of magnitude smaller than the current level of precision
in $^{38m}$K \cite{gorelov05} and $^{32}$Ar \cite{adelberger99}
and the anticipated goal in $^{35}$Ar \cite{beck03}.

The second experiment will use the Texas A\&M University TRAP (TAMUTRAP)
which is also a Penning trap system currently under construction \cite{mehlman13}.
The experimental program of TAMUTRAP includes
measurements of $a$ in a set of
super-allowed pure Fermi transitions with isospin $T=2$,
which are $\beta$-delayed
proton emitters, like $^{32}$Ar decay. The measuring principle relies on the
the broadening
of the delayed proton
energy spectrum, which is affected by the correlation between the $\beta$
particle and the neutrino \cite{adelberger99}.
The Penning trap setup employs an optimized length-to-radius ratio in the electrode
structure providing a 90~mm large inner radius \cite{mehlman13}.

The fact that the constraints on tensor couplings are considerably weaker
than those
on scalar ones (Figs.~\ref{fig:ReS-limNuclExp} and \ref{fig:ReT-limNuclExp})
has motivated a number of new experiments for precision
measurements in Gamow-Teller transitions, most of them focused on the
measurement of the $\beta\nu$ correlation coefficient in the decay of $^6$He.

A new measurement of $a$ has been carried
out at the Grand Acc\'el\'erateur National d'Ions Lourds (GANIL), Caen
in the $\beta$ decay of $^{6}$He \cite{flechard11}.
The ions were stored in a transparent Paul trap (LPCTrap)
\cite{rodriguez06} and the $\beta\nu$ correlation was deduced
from the time-of-flight spectrum of $^{6}$Li recoils detected in coincidence
with the $\beta$ particles \cite{flechard08}.
Three data production runs have been completed with this setup including a measurement
of the shake-off probability of the singly bound electron in the recoiling
ions following $\beta$ decay \cite{couratin12}.
A relative statistical error of 2\% has been obtained from the analysis of the first
run \cite{flechard11} with a systematic error of comparable size.

Since no limit on tensor couplings has been quoted in Ref.~\cite{flechard11} we 
review here the result in order
to further stress the sensitivity to the linear
contributions of the exotic couplings via the Fierz term.
The experimental result from Ref.~\cite{flechard11}
reads $\tilde{a}=-0.334(10)$ where the statistical and systematic errors have been added
in quadrature. For a Gamow-Teller transition, the expression of $b$ in Eq.~(\ref{eq:a-tilde})
is given in Eq.~(\ref{eq:bGT}).
Assuming first that the sensitivity to tensor couplings arises only via the Fierz
term, we then set $a \approx a_{SM} \approx -1/3$. The value
of the term $\langle m/E_e \rangle$ under the conditions of that experiment is
$\langle m/E_e \rangle = 0.20$~\footnote{For the purpose followed here,
the value of $\langle 1/E_e \rangle = \langle 1/(T_e + m_e) \rangle$ was
extracted from the spectrum of the measured kinetic energy, $T_e$,
without deconvoluting the detector response function.}.
Assuming $C_T = C^\prime_T$, the limit obtained from this measurement is then
\bea
\label{eq:CTCA-from-He6}
|C_T/C_A| < 0.13~~~~~~~\mbox{(90\% CL, first run LPCTrap)}.
\eea
If, in contrast, the contribution of the Fierz term is ignored, the limit
extracted from the quadratic contribution to $a$ would be $|C_T/C_A| < 0.23$ what
is
a factor of about 2 weaker. We see that, even if the sensitivity to the Fierz
term is here only $\langle m/E_e \rangle =0.20$, it remains
nevertheless dominant in the extraction of constraints on exotic couplings
from the expression of $\tilde{a}$.

A new measurement has been carried out with the LPCTrap at GANIL,
resulting in a relative statistical
precision $\delta a/a = 4.5\times10^{-3}$ at 1$\sigma$ \cite{ban13}.
Assuming that the final systematic error will be of comparable magnitude
than the statistical one, the result will provide the following bound
\beq
\label{eq:CT-LPC-final}
\left|{\rm Re}\left( \frac{C_T + C_T^\prime}{C_A} \right)\right| \lesssim
0.052~~~~~~~\mbox{(90\% CL)}~.
\eeq
This will finally improve the fifty-year old measurement of $a$ in $^6$He decay
\cite{johnson63}, but will still be a factor of 2.5 away from the current limit
obtained from $P_F/P_{GT}$, Eq.~(\ref{eq:b-PFPGT}). 

A high intensity gaseous source of $^{6}$He
has been developed at the Center for Experimental Nuclear Physics and Astrophysics
(CENPA)
in Seattle \cite{knecht11}. The measured extracted rate of atomic  $^{6}$He
available at a low background experimental area was about $~$10$^{9}$ atoms/s.
The source has been used for a high precision measurement of the $^{6}$He half-life
\cite{knecht12a}, with the atoms confined in a
cylindrical storage volume. The physics plans include a measurement of $a$
and of the Fierz interference term \cite{knecht11}.
The measurement of $a$ will be performed with atoms
confined in a MOT \cite{knecht12b} and by detecting the $\beta$ particles in
coincidence with the recoil ions. The goal of this project is to reach
the precision level of 1\% in a first phase and reach an ultimate precision
of 0.1\% after
possible improvements of the setup \cite{knecht12b}. 
Such a total precision would entail the following bound on tensor
interactions
\beq
\left|{\rm Re}\left( \frac{C_T + C_T^\prime}{C_A} \right)\right| \lesssim
0.0082~~~~~~~\mbox{(90\% CL)}~.
\eeq
This would improve the bound from nuclear and neutron decays,
Eq.~\eqref{eq:CT-val-globFit}, by a factor of about 2. The current bound from
radiative pion decay given in Eq.~\eqref{eq:tradpi} would still be slightly
stronger though.

An electrostatic ion beam trap is currently being built at
the Weizmann Institute of Science, in Rehovot \cite{aviv12}. The program around
this trap
includes a measurement of $a$ in $^6$He decay.
The ions will be trapped by mirror potentials, and the $^{6}$Li recoil ions
and the $\beta$ particles will be detected in coincidence in a field free
region \cite{aviv12}. The project envisages to take advantage of the high
yields of radioactive nuclei that will become available at the Soreq Applied
Research Accelerator Facility \cite{hirsh12}.

The $\beta$-$\alpha$-$\alpha$ correlation from $^8$Li decay has since long
been known to offer an attractive scheme in order to circumvent the detection
of the recoil nucleus \cite{christy47} although early results were not selective
enough to distinguish between different interaction types. After establishing
that the decay was dominated by a Gamow-Teller transition, a first limit on
a possible tensor contribution has been provided \cite{barnes58} from the
measurement of the $\alpha$ particle momenta. An additional advantage of 
this correlation in $^8$Li decay
is that, when the $\alpha$ and $\beta$ particles are detected along the same
direction, the sensitivity to the quadratic contributions of the tensor
interaction is enhanced by a factor of 3 \cite{barnes58} as compared to
a direct measurement of $a$ in a Gamow-Teller transition.
This decay has recently been reconsidered at Argonne National Laboratory using
ions stored in a longitudinal Paul trap \cite{scielzo12}. The $\alpha$
and $\beta$ particles are detected in coincidence
and the shape of the $\alpha$ particle energy
shift distribution is analyzed. The experiment has obtained a first result
\cite{li13} corresponding to a limit $|C_T/C_A|^2 < 0.026$ at 90\% CL.
It is expected that, with an upgraded detection system, the limit on $|C_T/C_A|^2$
can be improved by an order of magnitude \cite{li13}. This would then result
in a limit on the tensor couplings a factor of about 2 larger than
given in Eq.(\ref{eq:CT-LPC-final}).
Note that, since the $\beta$ particle end-point energy is rather large
in $^8$Li decay (16.1 MeV) the sensitivity to the Fierz term 
via the factor $\langle m/E_e \rangle$ is 3.3 times smaller than
in $^6$He decay.

Several experiments are currently ongoing aiming at improved measurements of
$a$ in neutron decay \cite{simson09,wietfeldt09,pocanic09}.
Such efforts are primarily motivated by the extraction of the ratio $g_A/g_V$
using another observable than the $\beta$ asymmetry parameter, $A$, in order to
determine the $V_{ud}$ matrix element from the neutron lifetime.
The anticipated precision goals on $\delta a/a$ of these experiments are of
<1\% for $a$CORN \cite{wietfeldt09}
<0.5\% for $a$SPECT \cite{simson09} and 0.1\% for Nab \cite{pocanic09}.
This last collaboration also aims a direct determination of the Fierz term
that we address here below.

In summary, the most precise experiments in Fermi transitions have reached
a level of
of few $10^{-3}$ at $1\sigma$ for the measurements of $a$
\cite{adelberger99,gorelov05}.
Comparable levels of precision have been anticipated by new projects
\cite{beck03} and it is reasonable to expect that these will soon be achieved.
It appears challenging that the new round of measurements of $a$ would be able to
improve the bounds on scalar couplings set by the Fierz term extracted from the
${\mathcal F}t$-values in pure Fermi transitions, Eq.~(\ref{eq:bF}).
However, this constraint has a stronger dependence to
theoretical uncertainties related to nuclear structure and radiative
corrections than those obtained from correlation coefficients.
Since the nuclear matrix elements cancel
to first order in the correlation coefficients, it is expected that nuclear
effects would manifest at the level of induced weak currents.
This leaves significant room for improvements in precision for measurements
of correlation coefficients.

For pure
Gamow-Teller transitions, considering the number of ongoing
experiments in $^6$He, $^8$Li and neutron decays
and the anticipated precision goals on a longer time
scale \cite{knecht12b,pocanic09},
it is expected that the next round of measurements of $a$
could reach a precision at the 0.1\% level.
This should have a significant impact to further constraint possible
tensor contributions.

%%%%%%%%%%%%%%%%%%%%%%%%%%%%%%%%%%%%%%%%%%%%%%%%%%%%%%%%%%%%%%%%%%%%%%%%%%%%%%%%%%%%%%%%%%%%
\subsection{The Fierz interference term, $b$}

The Fierz term can be accessed directly by
measurements of the shape of the $\beta$ energy spectrum. Such measurements
are very challenging, in particular because of instrumental
difficulties associated with the detection of $\beta$ particles at low energies
such as backscattering, out-scattering, detector dead-layers,
noise, etc.

The physics program of the CENPA group around
the high intensity
$^6$He source includes a measurement of $b$ in $^6$He decay \cite{knecht11}.
The technical details of
the detection system to be used for such a measurement and the sensitivity level
of the experiment have not yet been anticipated.

The problems associated with the scattering of $\beta$ particles in matter and
the limited precision of available data on low-energy electron scattering that is
required
for precise Monte-Carlo simulations, have motivated the development of a new and
compact $\beta$
spectrometer (miniBETA) by groups from Krakow and Leuven \cite{lojek09}. The system
is based on an ultra-light multi-wire drift chamber combined with energy sensitive
detectors. Precision measurements of the shape of $\beta$ energy spectra for transitions
with relatively low endpoint energies are being considered
\cite{severijns13a}.

The Fierz term has never been measured in neutron decay and there
are currently two projects aiming a direct determination from the
shape of the $\beta$ energy spectrum.

The Nab collaboration will use a cold neutron beam at
the Fundamental Neutron Physics Beamline of the Spallation Neutron Source
in Oak Ridge \cite{pocanic09}. The setup
uses a field-expansion magnetic spectrometer that
confines electrons and protons toward segmented Si detectors. 
The anticipated precision goal for the direct measurement of $b$
is $\delta b = 3\times 10^{-3}$ \cite{pocanic09} and is aimed to provide
an independent limit on exotic tensor couplings. 

The second project is the UCNb experiment at Los Alamos National Laboratory
\cite{hickerson13} and uses ultra-cold neutrons confined in a 4$\pi$ box made
of plastic scintillators. This experiment aims a precision of
$\delta b \approx 10^{-3}$ \cite{hickerson13}.
%
%%%%%%%%%%%%%%%%%%%%%%%%%%%%%%%%%%%%%%%%%%%%%%%%%%%%%%%%%%%%%%%%%%%%%%%%%%%%%%%%%%%%%%%%%%%%
\subsection{Time-reversal violating correlations}
A new measurement of the time-reversal violating triple correlation coefficient
$R$ has been carried out in $^8$Li decay at ISAC-TRIUMF \cite{murata11}.
The nuclear polarization is obtained by collinear
laser optical pumping of a low-energy beam which is then ionized an implanted 
on a Pt foil. The $\beta$ particle transverse
polarization is analyzed by Mott scattering on a lead foil and the particles
are tracked using planar wire chambers.
The precision goal of this experiment is to reach the level of final state
interactions which, as already indicated, are at $10^{-4}$ level for
this decay \cite{huber03}.
%
%%%%%%%%%%%%%%%%%%%%%%%%%%%%%%%%%%%%%%%%%%%%%%%%%%%%%%%%%%%%%%%%%%%%%%%%%%%%%%%%%%%%%%%%%%%%
\section{Summary and Outlook}
We have reviewed the status of the searches for
physics beyond the SM by precision measurements in nuclear and neutron $\beta$
decays. For the description of such processes at the quark-level, we
have used a model-independent EFT approach, assuming that the NP would emerge
at much higher energies, not only compared to those available in $\beta$ decays
but also those currently accessed or projected to be accessed at the LHC.
An attractive feature of this approach is that it provides a unified framework
to describe the effects of NP in nuclear, neutron and pion decays as well as
in collider physics.

It was shown that only three CP-conserving and three CP-violating NP couplings
can contribute linearly to the observables in $\beta$ decay, and so only those
couplings can be strongly probed with precision experiments at low energies.

We have then review the constrains on these six couplings from
precise measurements in nuclear and neutron decays, using the hadron-level
coefficients.
Except for the ${\mathcal F}t$-values deduced from pure Fermi transitions,
which combines results from a large set of experiments and for several
nuclear transitions,
we have selected the most precise results from single experiments. A more detailed analysis
would require a new global fit of all currently available data.
When analyzing the sensitivity of the
most stringent constraints on scalar and tensor couplings provided by measurements
of correlations or decay rates, we
found that the only relevant parameter for all constraints
is the Fierz interference term.
This term is linear in the couplings and contribute to the measured correlations
in most experimental conditions,
offering then the dominant sensitivity to those couplings.

With the most precise results, we have then moved to a quark-level description
in order to compare first with results obtained from other semi-leptonic
processes but also with results obtained from measurements at the LHC.
This illustrates the benefit of the EFT approach for the sensitivity comparison
between different hadronic probes, such as pions, nucleons, and nuclei.
For the CP-conserving coefficients, direct limits from $\beta$ decay appear to
be very competitive.
Next we discussed the interplay with results from the LHC and we stressed
the complementary of precision measurements in $\beta$ decay as probes of
new physics. An interesting competition with results from the LHC arises since
the effective scale probed by low energy experiments overlaps with the LHC reach.
The most attractive scenario of such interplay would be that in which a non-zero
result would be observed for an exotic effective coupling either at low energies
or in the collider searches. If a new particle were
found at the LHC, experiments in $\beta$ decays will play an important role in
disentangling the properties of the NP dynamics.
We also presented the projected sensitivities to be reached at LHC which are
important to orient new precision goals for measurements at low energies.

We have finally reviewed the current experimental efforts looking for signatures
of NP in measurements of correlations or spectrum shapes in nuclear and neutron
decays. The purpose was to project
the sensitivity level of those efforts and confront them with the future
LHC reach. We have stressed again the importance of the Fierz term in those
searches and gave simple quantitative illustrations in the
measurement of $a$ in Fermi and Gamow-Teller transitions. Although
recent measurements of the $\beta$ asymmetry parameter $A$ in Gamow-Teller
transitions have explicitly relied on the sole contribution of the Fierz term
in the search for tensor interactions,
the role this term does not appear to have been generally incorporated
for the optimization of current experiments and for the design of new projects.

The sensitivity goal imposed by future LHC reach is very challenging but possibly
within reach by next generation experiments, where precisions at the level of
$10^{-3}$ or below are needed on the Fierz term. For the comparison with the
SM predictions of the measured observables, accurate theoretical calculations
are fundamental. These require the inclusion of radiative corrections and recoil
order effects and, whenever necessary, of nuclear structure corrections. The
errors on these corrections set the next theoretical limit of sensitivity
for future measurements
and leave still a large window for improvements for experiments measuring
correlations or spectrum shapes in nuclear and neutron decays.

%%%%%%%%%%%%%%%%%%%%%%%%%%%%%%%%%%%%%%%%%%%%%%%%%%%%%%%%%%%%%%%%%%%%%%%%%%%%%%%%%%%%%%%%%%%%
\section{Acknowledgements}
We are grateful to V.~Cirigliano, X.~Fl\'echard, V.~Gudkov,
K. Minamisono,
N.~Severijns, M.~Sternberg and S.~Tulin for stimulating discussions
and for correspondence.
This work was supported in part by the U.S. National Science Foundation under
grant number PHY-11-02511, the DOE contract DE-FG02-08ER41531 and the
Wisconsin Alumni Research Foundation.

%%%%%%%%%%%%%%%%%%%%%%%%%%%%%%%%%%%%%%%%%%%%%%%%%%%%%%%%%%%%%%%%%%%%%%%%%%%%%%%%%%%%%%%%%%%%

%%%%%%%%%%%%%%%%%%%%%%%%%%%%%%%%%%%%%%%%%%%%%%%%%%%%%%%%%%%%%%%%%%%%%%%%%%%%%%%%%%%%%%%%%%%%
\end{document}